\documentclass[journal]{IEEEtran}
\ifCLASSINFOpdf
\else
\fi

\usepackage{amsmath}
\usepackage{amsfonts}
\usepackage{amssymb}
\usepackage{cite}
\usepackage{graphicx}
\usepackage{textcomp}
\usepackage{xcolor}
\usepackage{subfigure} 
\usepackage{url}
\usepackage{array}
\usepackage{multirow}
\usepackage{booktabs}
\usepackage{array}
\usepackage{tabularx}
\usepackage{graphicx}
\usepackage{booktabs}
\usepackage{xcolor}
\usepackage{pifont}
\usepackage{threeparttable}
\renewcommand{\arraystretch}{1.1}
\usepackage{booktabs}
\usepackage{tabularx}

\newcommand{\cmark}{\textcolor[rgb]{0.0,0.5,0.0}{\ding{51}}}  
\newcommand{\xmark}{\textcolor[rgb]{0.6,0.0,0.0}{\ding{55}}}  

\begin{document}

\title{ViPSN~2.0: A Reconfigurable Battery-free IoT Platform for Vibration Energy Harvesting}
\author{
    Xin~Li*,~\IEEEmembership{Member,~IEEE,}
    Mianxin~Xiao,~\IEEEmembership{Student~Member,~IEEE,}
    Xi~Shen,
    Jiaqing~Chu,
    Weifeng~Huang,~\IEEEmembership{Student~Member,~IEEE,}
    Jiashun~Li,
    Yaoyi~Li,
    Mingjing~Cai,
    Jiaming~Chen,
    Xinming~Zhang,
    Daxing~Zhang,
    Congsi~Wang,~\IEEEmembership{Senior~Member,~IEEE,}
    Hong~Tang,~\IEEEmembership{Student~Member,~IEEE,}
    Bao~Zhao,~\IEEEmembership{Member,~IEEE,}
    Qitao~Lu,~\IEEEmembership{Member,~IEEE,}
    Yilong~Wang, 
    Jianjun~Wang,
    Minyi~Xu,~\IEEEmembership{Member,~IEEE,}
    Shitong~Fang,
    Xuanyu~Huang,
    Chaoyang~Zhao,~\IEEEmembership{Member,~IEEE,}
    Zicheng~Liu,~\IEEEmembership{Member,~IEEE,}
    Yaowen~Yang,~\IEEEmembership{Senior~Member,~IEEE,}
    Guobiao~Hu,~\IEEEmembership{Member,~IEEE,}
    Junrui~Liang*,~\IEEEmembership{Senior~Member,~IEEE},
    and~Wei-Hsin~Liao*,~\IEEEmembership{Senior~Member,~IEEE}
	\thanks{
    This work was supported in part by the Guangdong Basic and Applied Basic Research Foundation (Grant No. 2025A1515011342), The Chinese University of Hong Kong (Grant No. 3134164 and 4055178), the China Postdoctoral Science Foundation (Grant Nos. 2024M761632 and BX20240183), the Shenzhen Science and Technology Program (Grant No. KQTD20240729102211015), and the National Natural Science Foundation of China under Grant (Grant Nos. 62271319 and U21B2002). (\textit{Co-first authors: Xin Li and Mianxin Xiao}; \textit{Corresponding authors: Xin Li, Junrui Liang, and Wei-Hsin Liao}.)
    
    Xin Li is with the Guangzhou Institute of Technology, Xidian University, China, also with the Department of Mechanical and Automation Engineering, The Chinese University of Hong Kong, Hong Kong (e-mail: lixin01@xidian.edu.cn).
    Junrui Liang is with the the School of Information Science and Technology, ShanghaiTech University, Shanghai, China (e-mail: liangjr@shanghaitech.edu.cn).
    Wei-Hsin Liao is with the Department of Mechanical and Automation Engineering, The Chinese University of Hong Kong, Hong Kong, China (e-mail: whliao@cuhk.edu.hk).
    
    Mianxin Xiao, Xi Shen, Jiaqing Chu, Weifeng Huang, Jiashun Li, Yaoyi Li, Mingjing Cai, Jiaming Chen, Xinming Zhang, Daxing Zhang, and Congsi Wang are with the Guangzhou Institute of Technology, Xidian University, Guangzhou, China.
    Hong Tang is with the Department of Computer Science and Engineering at Shanghai Jiao Tong University, Shanghai, China.
    Bao Zhao is with the Department of Civil and Environmental Engineering, Hong Kong Polytechnic University, Hung Hom, Kowloon, Hong Kong, China.
    Qitao Lu is with the Department of Mechanical and Automation Engineering, The Chinese University of Hong Kong, Hong Kong, China.
    Yilong Wang is with the School of Astronautics, Harbin Institute of Technology, Harbin, China.
    Jianjun Wang is with the Department of Applied Mechanics, University of Science and Technology Beijing, Beijing, China.
    Minyi Xu is with the Marine Engineering College, Dalian Maritime University, Dalian, China. 
    Shitong Fang is with the College of Mechatronics and Control Engineering, Shenzhen University, Shenzhen, China.
    Xuanyu Huang is with the Tsinghua Shenzhen International Graduate School, Tsinghua University, Shenzhen, China, also with the Institute of Superlubricity Technology, Research Institute of Tsinghua University in Shenzhen, Shenzhen, China. 
    Chaoyang Zhao, Zicheng Liu, and Yaowen Yang are with the with the School of Civil and Environmental Engineering, Nanyang Technological University, Singapore.
    Guobiao Hu is with the Thrust of Internet of Things, The Hong Kong University of Science and Technology (Guangzhou), Guangzhou, China.
    }
    }

\markboth{Manuscript submitted to \textit{IEEE INTERNET OF THINGS JOURNAL}}%
{Xin Li \MakeLowercase{\textit{et al.}}: ViPSN~2.0}

\maketitle
\begin{abstract}
Vibration energy harvesting is a promising solution for powering battery-free IoT systems; however, the instability of ambient vibrations presents significant challenges, such as limited harvested energy, intermittent power supply, and poor adaptability to various applications. To address these challenges, this paper proposes ViPSN2.0, a modular and reconfigurable IoT platform that supports multiple vibration energy harvesters (piezoelectric, electromagnetic, and triboelectric) and accommodates sensing tasks with varying application requirements through standardized hot-swappable interfaces. ViPSN~2.0 incorporates an energy-indication power management framework tailored to various application demands, including light-duty discrete sampling, heavy-duty high-power sensing, and complex-duty streaming tasks, thereby effectively managing fluctuating energy availability. The platform's versatility and robustness are validated through three representative applications: ViPSN-Beacon, enabling ultra-low-power wireless beacon transmission from a single transient fingertip press; ViPSN-LoRa, supporting high-power, long-range wireless communication powered by wave vibrations in actual marine environments; and ViPSN-Cam, enabling intermittent image capture and wireless transfer. Experimental results demonstrate that ViPSN~2.0 can reliably meet a wide range of requirements in practical battery-free IoT deployments under energy-constrained conditions.
\end{abstract}

\begin{IEEEkeywords}
Vibration energy harvesting, Battery-free IoT, Piezoelectric, Magnetoelectric, Triboelectric
\end{IEEEkeywords}
\IEEEpeerreviewmaketitle

\section{Introduction}
\label{Introduction}

\begin{figure} [t]
	\centering
	\includegraphics[width=\columnwidth]{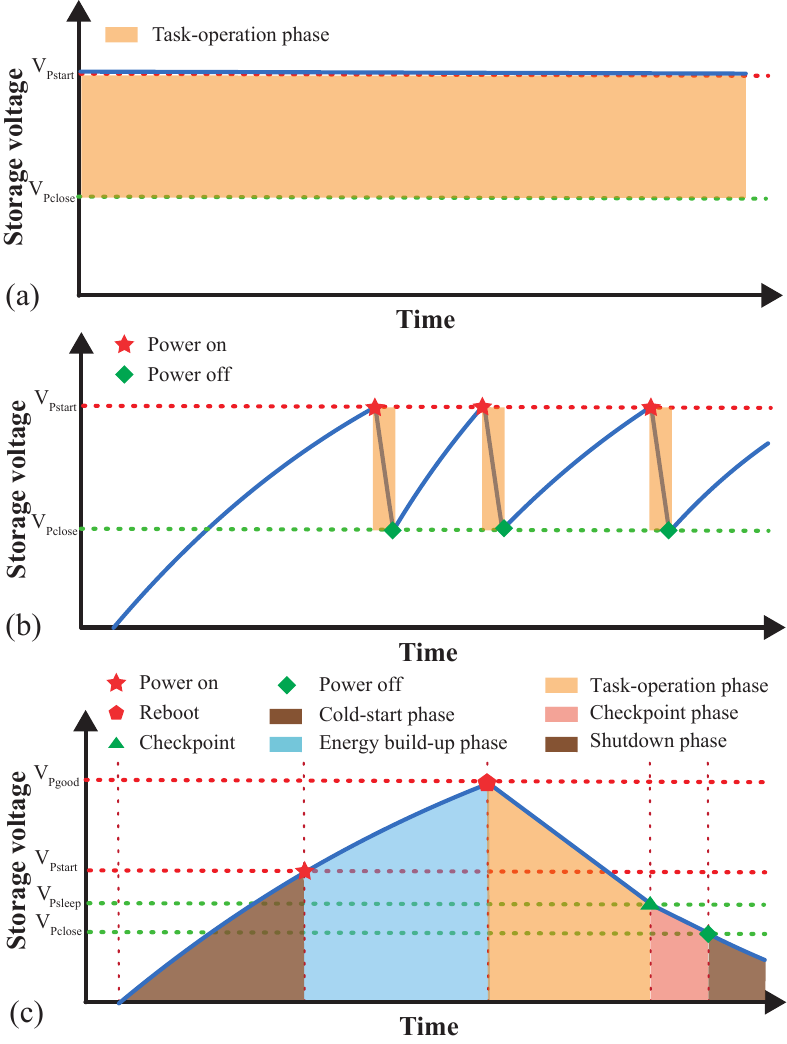}
	\caption{
        Comparison of operating modes for battery-powered and vibration-powered IoT systems. 
        (a) Continuous and stable operation of a battery-powered system;
        (b) Intermittent operation of a vibration-powered system; 
        (c) Energy-driven partitioning of system states in a vibration-powered IoT device, including cold start, energy build-up, task operation, checkpoint, and shutdown phases.
        }
	\label{fig_operation_comp}
\end{figure}

Vibration energy harvesting (VEH) has emerged as a promising technology for enabling battery-free Internet of Things (IoT) systems. Unlike solar and radio-frequency (RF) energy harvesting, VEH is not constrained by sunlight availability or spatial distribution of RF sources, enabling it particularly suitable for diverse mechanical vibration-rich environments, such as industrial machinery, infrastructure monitoring, human motion tracking, and wildlife sensing~\cite{gorlatova2015movers}.

However, ambient vibrations are typically sparse, highly intermittent, and unpredictable, which makes efficient energy harvesting and utilization challenging \cite{Anton2007, Tang2010-zd, Harne2013-dv, Safaei2019}. As illustrated in Fig.~\ref{fig_operation_comp}, battery-powered IoT systems benefit from stable and abundant energy, enabling continuous sensing, computation, and wireless communication. In contrast, vibration-powered IoT systems lack reliable and continuous power sources due to environmental constraints, inevitably experiencing frequent power interruptions and intermittent operation cycles\cite{lucia2017intermittent}. Consequently, vibration-powered IoT systems must dynamically adapt to fluctuating energy availability, maximizing quality-of-service (QoS) within constrained and variable energy conditions.

\textbf{Power Limitation:} 
Due to the inherent size constraints of IoT devices, typical vibration energy harvesters achieve power densities of only 5--60~$\mu$W/cm$^3$, significantly below the power requirements (often exceeding 10 mW) of IoT sensors and wireless transceivers~\cite{Ottman2003-nx,Loong2018-nc}. Despite advances in nonlinear transducer designs~\cite{tao8281027} and sophisticated power-conditioning circuits~\cite{Ottman2002-xd,Liang2017-ol}, practical VEH outputs remain insufficient for continuous IoT operation. Energy buffering via supercapacitors can support brief high-power tasks, but extended recharge intervals compromise system responsiveness~\cite{orfei2016vibrations}. IoT-focused strategies primarily reduce computational and communication overhead, typically sacrificing QoS to match limited harvested energy~\cite{brenes2020maximum,Liu2021-pv}. However, due to the variability of vibration sources and IoT workloads, these approaches—either solely increasing harvested power or solely reducing energy consumption—cannot achieve stable power equilibrium as in battery-powered systems.

\textbf{Energy Fluctuation:} 
Ambient vibrations exhibit highly dynamic and unpredictable excitation patterns, ranging from regular harmonic vibrations in industrial machines to transient, irregular energy bursts from human motion~\cite{gao2019macro,jia2020review}. These variations frequently cause mismatches between energy availability and periodic or event-driven IoT workloads, leading to unavoidable power interruptions and inefficient energy utilization. Consequently, as shown in Fig. \ref{fig_operation_comp}(b), VEH-powered IoT systems can only operate within intermittent and discrete energy blocks, adapting their sensing, computation, and communication activities to available energy~\cite{hester2017future}. 
Effectively managing computations under such intermittent energy conditions remains a fundamental challenge for VEH-based IoT systems.

\textbf{Intermittent Execution:} Frequent power interruptions force vibration-powered IoT nodes into repeated reboot cycles, disrupting computations, causing data loss, and preventing the completion of complex or long-duration tasks~\cite{ransford2011mementos,colin2016chain}. In distributed sensing scenarios, intermittent execution further complicates synchronization and communication, as simultaneous active periods among nodes cannot be guaranteed, significantly affecting system reliability and scalability~\cite{nardello2019camaroptera,de2020reliable,sandhu2021task}.

\textbf{Multidisciplinary Collaboration:} Unlike commercially scaled solar panels or RF antennas, vibration energy harvesters require customized mechanical structures tailored to specific vibration modes (e.g., harmonic, intermittent, or impact-driven)\cite{cai2020recent}. However, current research often isolates mechanical design, electrical conversion circuits, and IoT system integration, neglecting critical interdependencies among these domains\cite{magno2016kinetic,li2018trinity}. Consequently, existing VEH-based IoT solutions remain limited mostly to simple, low-rate applications, such as temperature or humidity monitoring, restricting their practical utility compared to more advanced solar- or RF-powered IoT deployments~\cite{afanasov2020battery,de2020battery}. 
Addressing these interdisciplinary challenges through holistic design and system-level integration is essential for unlocking the full potential of VEH-based IoT systems in practical deployments.

\begin{table*}[t]
    \centering
    \footnotesize
    \caption{Representative Examples of Application Evolution from ViPSN~1.0 to ViPSN~2.0}
    \label{table_vipsn1_evolution}
    \begin{tabularx}{\textwidth}{p{0.57cm} p{1.7cm} p{1.2cm} p{2.7cm} p{0.8cm} p{1.6cm} p{1.6cm} X}
        \toprule
        \textbf{Year} & \textbf{User (Ref.)} & \multicolumn{2}{c}{\textbf{EEU}} & \textbf{EMU} & \textbf{EUU} & \textbf{Peripheral} & \textbf{Application Field} \\
        \cmidrule(lr){3-4}
        &  & \textbf{Transducer} & \textbf{Source} & & & & \\
        \midrule
        2020 & K. Ren~\cite{zhang2021wind} & TENG & Wind-induced vibration & Buck & BLE Beacon & Temp. & Environmental sensing, smart buildings \\
        2022 & Y. Yang~\cite{hu2022triboelectric} & TENG & Human motion & Buck & BLE Beacon & Temp. & Wearable devices, human health monitoring \\
        2022 & J. Wang~\cite{wang2022piezoelectric} & PZT & Railway vibration & Buck & BLE Beacon & Temp. & Railway operation, structural health monitoring \\
        2023 & C. Zhao~\cite{zhao2023wide} & TENG & Shaker vibration & Buck & BLE Beacon & Temp. & Environmental sensing \\
        2024 & G. Hu~\cite{wang2024rolling} & TENG & Ocean wave vibration & Buck & LoRa & TDS & Marine environmental monitoring \\
        2024 & K. Chen~\cite{chen2024design} & PZT & Shaker vibration & Buck & LoRa & Temp. & Structural health monitoring \\
        2024 & L. Dong~\cite{dong2025advanced} & TENG & Wind-induced vibration & Buck & BLE Beacon & Temp. & Railway tunnel structure monitoring \\
        2024 & Q. Lu~\cite{lu2024ultra} & EMG & Human motion & Boost & BLE Beacon & Temp. & Wearable devices, human health monitoring \\
        2024 & S. Fang~\cite{fang2024high} & PZT & Turbine vibration & Buck & LoRa & Temp., Hum. & Intelligent operation, structural monitoring \\
        2025 & Z. Liu~\cite{liu2025electromechanical} & TENG & Shaker vibration & Buck & BLE Beacon & Temp. & Structural health monitoring \\
        2025 & Y. Wang~\cite{wang2025mutualistic} & PZT & Vehicle vibration & Linear & BLE Beacon & Temp. & Vehicle-to-everything, intelligent transportation \\
        2025 & Y. Li~\cite{li2025heat} & TENG & Temperature difference-induced vibration & Buck & BLE Beacon & Temp. & Environmental sensing, structural monitoring \\
        \midrule
        \textbf{Year} & \textbf{Institution} & \textbf{Transducer} & \textbf{Source} & \textbf{EMU} & \textbf{EUU} & \textbf{Peripheral} & \textbf{Application Field} \\
        \midrule
        2021 & NPU, CN & PZT & Shaker vibration & Buck & BLE Beacon & Temp. & Environmental sensing \\
        2021 & BJTU, CN & PZT & Railway vibration & Buck & BLE Beacon & Temp. & Railway operation, structural health monitoring \\
        2021 & HUST, CN & PZT, TENG & Wind-induced vibration & Buck & BLE Beacon & Temp. & Environmental monitoring \\
        2021 & ZJNU, CN & PZT & Machine-induced vibration & Buck & BLE Beacon & Temp. & Equipment monitoring \\
        2021 & ZZU, CN & PZT & Wind-induced vibration & Buck & BLE Beacon & Temp. & Environmental monitoring \\
        2022 & ETH Zürich & PZT & Machine-induced vibration & Buck & BLE Beacon & -- & Metamaterial structure sensing \\
        2023 & DUT, CN & TENG & Ocean wave vibration & Buck & BLE Beacon & Temp. & Marine environmental monitoring \\
        2023 & UNNC, CN & EMG & Machine-induced vibration & Boost & BLE Beacon & -- & Equipment monitoring \\
        2023 & BJUT, CN & PZT & -- & Buck & BLE Beacon & Temp. & Environmental monitoring \\
        2023 & XJTU, CN & PZT & -- & Buck & BLE Beacon & Temp. & Structural health monitoring \\
        2023 & CSU, CN & PZT & Railway vibration & Buck & BLE Beacon & -- & Railway operation, structural health monitoring \\
        2023 & HIT, CN & PZT & Machine-induced vibration & Buck & BLE Beacon & -- & Structural health monitoring \\
        2023 & BINN, CN & TENG & Wind-induced vibration & Buck & BLE Beacon & -- & Environmental monitoring \\
        2024 & UoA, NZ & EMG & Wind-induced vibration & Boost & Lora & Temp., Hum. & Bridge structure monitoring, environmental monitoring \\
        2024 & THU, CN & EMG & Power line swing & Boost & BLE UART & Acc., Temp. & Power transmission online monitoring \\
        2024 & NUEC, CN & EMG & Engine vibration & Linear & BLE Beacon & Temp. & Equipment operation and maintenance monitoring \\
        2025 & CQU, CN & TENG & -- & Buck & BLE Beacon & -- & Environmental monitoring, structural health monitoring \\
        2025 & HKUST(GZ), CN & EMG & Vehicle-induced vibration & Boost & BLE UART & Cam. & Intelligent transportation management, safety and emergency \\
        \bottomrule
    \end{tabularx}
\end{table*}


To address these critical challenges, we introduce ViPSN~2.0, a modular and reconfigurable IoT platform specifically designed for VEH-powered battery-free applications. ViPSN~2.0 integrates multiple vibration transducers, including piezoelectric (PZT), electromagnetic (EMG), and triboelectric (TENG) harvesters, via standardized hot-swappable interfaces, enabling rapid deployment and adaptation across diverse vibration-rich environments. The platform incorporates an energy indication-based power management framework that dynamically allocates harvested energy to sensing, computation, and communication tasks based on instantaneous energy availability and application-specific QoS requirements. 

We demonstrate the practicality and effectiveness of ViPSN~2.0 through three representative applications:
\begin{itemize}
    \item \textbf{ViPSN-Beacon} representing the \textit{light-duty ultra-low-energy} applications. It enables wireless beacon transmissions powered entirely by only one transient fingertip press excitation.
    \item \textbf{ViPSN-LoRa} representing the \textit{heavy-duty long-distance communication} applications. It supports reliable kilometer-scale wireless communication powered by wave-induced vibrations in marine environments.
    \item \textbf{ViPSN-Cam} representing the \textit{complex-duty data streaming} applications. It facilitates streaming image capture and wireless transfer under intermittent vibration conditions.
\end{itemize}

\begin{table*}[!t]
    \centering
    \caption{Comparison of Platforms Supporting Various Transducer Types and Excitation Methods}
    \label{table_platform_comparison}
    \begin{tabularx}{\textwidth}{p{0.57cm} p{2cm} p{0.7cm} p{0.7cm} p{0.7cm} p{1.1cm} p{1.1cm} p{1cm} p{1.2cm} p{2.42cm} X}
        \toprule
        \textbf{Year} & \textbf{Platform} 
        & \multicolumn{3}{c}{\textbf{Transducer Types}} 
        & \multicolumn{3}{c}{\textbf{Excitation Types}}
        & \textbf{Discrete Sampling}
        & \textbf{Streaming Sampling}
        & \textbf{Remarks} \\
        \cmidrule(lr){3-5}\cmidrule(lr){6-8}
        & & \textbf{PEH} & \textbf{EMG} & \textbf{TENG}
          & \textbf{Continuous} & \textbf{Intermittent} & \textbf{Transient}
          & & & \\
        \midrule
        2006 & WISP~\cite{smith2006wirelessly} & \xmark & \xmark & \xmark & \xmark & \xmark & \xmark & \cmark & \xmark & RF only \\
        2008 & NeuralWISP~\cite{holleman2008neuralwisp} & \xmark & \xmark & \xmark & \xmark & \xmark & \xmark & \cmark & \xmark & RF only \\
        2016 & WISPCam~\cite{naderiparizi2016wispcam} & \xmark & \xmark & \xmark & \xmark & \xmark & \xmark & \xmark & \cmark\ (Cam.) & RF only \\
        2017 & Flicker~\cite{hester2017flicker} & \cmark & \xmark & \xmark & \cmark & \xmark & \xmark & \cmark & \cmark\ (Acc.) & No VEH tests \\
        2018 & Capybara~\cite{colin2018reconfigurable} & \xmark & \xmark & \xmark & \xmark & \xmark & \xmark & \cmark & \xmark & Solar only \\
        2019 & Shepherd~\cite{geissdoerfer2019shepherd} & \xmark & \xmark & \xmark & \xmark & \xmark & \xmark & \cmark & \xmark & Solar only \\
        2019 & AsTAR~\cite{yang2019astar} & \xmark & \xmark & \xmark & \xmark & \xmark & \xmark & \cmark & \cmark\ (Acc.) & Solar only \\
        2022 & Tartan~\cite{denby2022tartan} & \xmark & \xmark & \xmark & \xmark & \xmark & \xmark & \cmark & \cmark\ (Acc.) & Solar only \\
        2022 & MakeCode~\cite{kraemer2022battery} & \cmark & \xmark & \xmark & \cmark & \xmark & \xmark & \cmark & \cmark\ (Acc.) & No VEH tests \\
        2022 & Protean~\cite{bakar2022protean} & \xmark & \xmark & \xmark & \xmark & \cmark & \xmark & \cmark & \cmark\ (Cam., Aud., Acc.) & Solar only \\
        2024 & Riotee~\cite{geissdoerfer2024riotee} & \cmark & \xmark & \xmark & \cmark & \xmark & \xmark & \cmark & \cmark\ (Acc., Mic.) & No VEH tests \\
        2020 & ViPSN~1.0~\cite{li2020vipsn} & \cmark & \xmark & \xmark & \cmark & \cmark & \cmark & \cmark & \xmark & PZT only \\
        2025 & ViPSN~2.0 [This Work] & \cmark & \cmark & \cmark & \cmark & \cmark & \cmark & \cmark & \cmark\ (Acc., Cam.) &  PZT, EMG, TENG \\
        \bottomrule
    \end{tabularx}
\end{table*}

\section{Past Experience and Related Work}

ViPSN~2.0 builds upon the foundation of ViPSN 1.0~\cite{li2020vipsn}, the first open-source IoT platform explicitly designed for vibration-powered applications. 
ViPSN~1.0 initially supported only PZT harvesters, and practical deployments revealed several critical limitations, including the absence of standardized multi-transducer interfaces, limited scalability for supporting multiple peripheral modules, lack of adaptive energy management, and an inflexible power management system.

During the transition to ViPSN~2.0, the platform has supported over 50 research and educational projects spanning multiple fields, demonstrating its versatility and scalability. As shown in Table~\ref{table_vipsn1_evolution}, Wang et al.~\cite{zhang2021wind} utilized PZT harvesters to collect railway vibrations, enabling structural health monitoring; Lu et al.~\cite{lu2024ultra} integrated EMG harvesters into wearable devices powered by human motion, enabling continuous health monitoring; and Hu et al.~\cite{wang2024rolling} integrated TENG harvesters with a long-range wireless module to develop an ocean wave-powered IoT system for remote water quality monitoring. These projects highlight how ViPSN has progressively expanded its scope, evolving from a PZT-only framework to a platform supporting various energy harvesters and application scenarios.

Meanwhile, battery-free IoT systems powered by RF and solar energy harvesting have significantly advanced, enabling complex sensing tasks such as wireless image streaming~\cite{naderiparizi2016wispcam} and audio capture~\cite{bakar2022protean}. However, as shown in Table~\ref{table_platform_comparison}, most existing platforms primarily target RF or solar harvesting. Widely used platforms such as WISP~\cite{smith2006wirelessly}, NeuralWISP~\cite{holleman2008neuralwisp}, and WISPCam~\cite{naderiparizi2016wispcam} exclusively support RF harvesting, lacking specialized mechanical designs and dedicated VEH circuit optimizations. Even general-purpose platforms claiming multi-harvester compatibility, such as Flicker~\cite{hester2017flicker} and Riotee~\cite{geissdoerfer2024riotee}, provide limited or no targeted evaluations for VEH scenarios, thus severely limiting their practical applicability in vibration-rich environments.

Additionally, dedicated VEH-powered IoT platforms remain relatively scarce and often exhibit several practical limitations. For example, Trinity~\cite{li2018trinity}, specifically developed for deployment in wind-induced vibration scenarios in heating, ventilation, and air conditioning (HVAC) systems, relies on large energy storage elements, resulting in extended cold-start delays. Additionally, its tightly integrated hardware architecture provides limited flexibility when adapting to varying vibration conditions and diverse application requirements. The inherently intermittent and unpredictable nature of vibration energy further complicates system operation, often causing computational interruptions, data inconsistency, and degraded QoS~\cite{hester2019batteries, liang2021kinetic}. 

Motivated by these practical experiences and identified limitations, ViPSN~2.0 systematically addresses these gaps through standardized multi-transducer interfaces, multiple power solution frameworks, and enhanced modularity and usability. These improvements aim to facilitate rapid prototyping, flexible system integration, and reliable operation across diverse vibration-powered IoT deployments.

\begin{figure}[t]
    \centering
    \includegraphics[width=1\columnwidth]{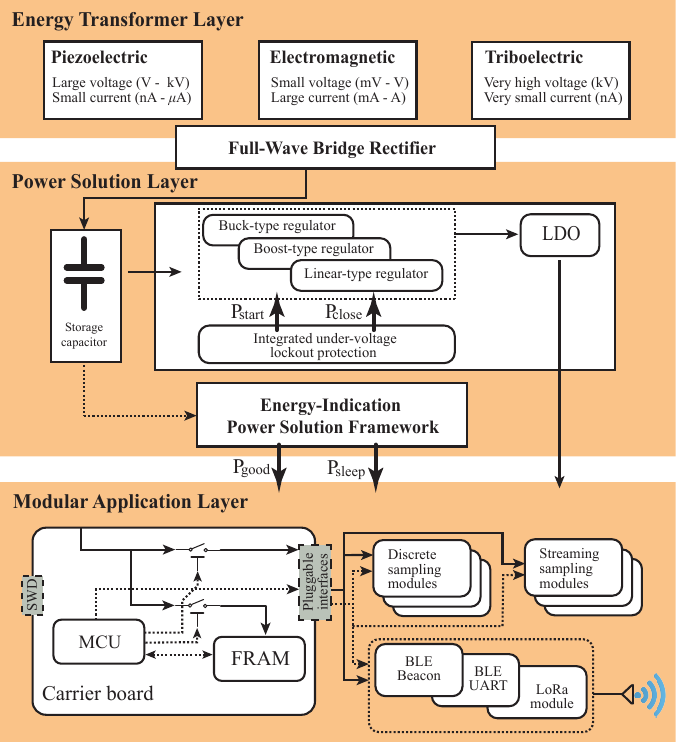 }
    \caption{System architecture of ViPSN~2.0, consisting of the energy transformation layer, power solution layer, and modular application layer.}
    \label{fig:fig_system_hardware_frame}
\end{figure}

\section{Platform Overview}
We designed and implemented ViPSN~2.0 as an open, modular platform for IoT developers targeting battery-free VEH-powered applications. The primary objectives of ViPSN~2.0 are: 
1) providing flexible hardware support for multiple VEH technologies, including PZT, EMG, and TENG, and various ambient excitations, including continuous, intermittent, and transient vibrations; 
2) offering robust and configurable power solutions capable of handling tasks ranging from discrete sampling to complex streaming data acquisition, with both short- and long-range wireless transmission; 
3) improving usability and enabling rapid prototyping through modular hardware and standardized software interfaces. 
Rather than a monolithic design, ViPSN~2.0 adopts interchangeable hardware modules and adaptive power solutions, allowing developers to customize system configurations according to specific application requirements and vibration conditions. Fig.~\ref{fig:fig_system_hardware_frame} illustrates the overall system architecture, and the following subsections detail each design component layer.

\subsection{Energy Transformer Layer}
The energy transformer layer supports interchangeable VEH transducers, each optimized for specific vibration conditions. PZT transducers provide high-voltage outputs (V to kV) and low-current levels (nA to $\mu$A), ideal for intermittent or transient forces like human motion or structural impacts. EMG transducers generate low-voltage (mV to V) and high-current outputs (mA to A), making them suitable for continuous vibrations like rotating machinery, enabling energy harvesting without mechanical contact. TENG transducers generate very high voltages (kV) at very low currents (nA), effectively capturing irregular vibrations caused by minimal displacement, including airflow or ambient structural vibrations. This layered design maximizes compatibility with diverse real-world vibration sources, thereby enhancing overall system flexibility and applicability.

\subsection{Power Solution Layer}  
The power management system integrates linear-, boost-, and buck-type regulators to support a wide range of input and output voltages. It features an energy management system with dynamically adjustable thresholds and four key operating signals:
\begin{itemize}
    \item $\mathrm{P_{start}}$: Indicates system startup.
    \item $\mathrm{P_{good}}$: Signals that the storage capacitor has accumulated sufficient energy to support the system's most energy-intensive atomic operation.
    \item $\mathrm{P_{sleep}}$: Warns that stored energy is insufficient for safe operation and initiates a transition to a low-power state.
    \item $\mathrm{P_{close}}$: Indicates system shutdown due to energy depletion.
\end{itemize}

As shown in Fig.~\ref{fig_operation_comp}(c), these signals divide the system's energy state into multiple functional phases, enabling explicit and fine-grained energy management throughout the entire power supply chain.
\begin{itemize}
    \item Cold-Start Phase ($0\,\mathrm{V}$ to $V_\text{Pstart}$): 
    Represents the initial energy accumulation from zero to the minimal energy level required for system startup. The system remains inactive during this phase until sufficient energy is accumulated.

    \item Energy Build-Up Phase ($V_\text{Pstart}$ to $V_\text{Pgood}$): 
    Introduced by the $\mathrm{P_{good}}$ signal, this additional phase allows further accumulation of energy beyond initial startup requirements, effectively extending the available energy block to reliably execute subsequent energy-intensive atomic operations.

    \item Task-Operation Phase ($V_\text{Pgood}$ to $V_\text{Psleep}$): 
    Provides sufficient stored energy for normal system operations, including sensing, computation, data storage, and wireless communication tasks.

    \item Checkpoint Phase ($V_\text{Psleep}$ to $V_\text{Pclose}$): 
    Reserved explicitly for emergency checkpointing operations, enabling the system to safely save critical state information (e.g., registers, stack, variables) into non-volatile memory, ensuring consistent recovery upon subsequent energy restoration.

    \item Shutdown Phase (below $V_\text{Pclose}$): 
    Indicates depletion of stored energy, forcing the system to enter a shutdown state until energy levels are replenished to reach the restart threshold.
\end{itemize}

By standardizing these energy signals and clearly defining their corresponding operational phases, the power solution layer establishes a flexible and modular foundation for implementing various power management strategies. This design enables ViPSN~2.0 to achieve explicit energy-level awareness and seamless integration of configurable power solutions, thereby supporting reliable execution and efficient energy utilization in vibration-powered IoT applications.

\subsection{Modular Application Layer}
The modular application layer is designed to accommodate the heterogeneous requirements of IoT applications by providing a standardized, plug-and-play interface for diverse sensor and communication modules. This structure supports a wide spectrum of workloads, ranging from discrete sampling---the periodic acquisition of data at fixed intervals, typically used in low-frequency sensing tasks such as temperature, humidity, or pressure monitoring---to streaming sampling, which involves the continuous, high-rate collection of sensor data required in more demanding scenarios, such as inertial measurement and vision-based applications.

To address varying data transmission needs, the platform integrates both short-range communication (for energy-efficient local data exchange, typically within 100 meters) and long-range transmission capabilities (supporting deployments over several kilometers), thus enabling flexible adaptation to different deployment scales. The standardized module interface facilitates rapid reconfiguration and seamless integration of heterogeneous sensors and communication modules, supporting scalable and reconfigurable application development. Overall, this modular design ensures that the platform can efficiently support light-duty tasks (infrequent, low-power operations such as periodic temperature measurements), heavy-duty tasks (high-power operations, for example, wireless transmissions over kilometer-scale distances), and complex-duty tasks (coordinated execution of advanced sensing, computation, and communication activities, such as image capture, with dynamic workload patterns), across varied IoT environments.

\begin{figure*}[t]
    \centering
    \includegraphics[width=2\columnwidth]{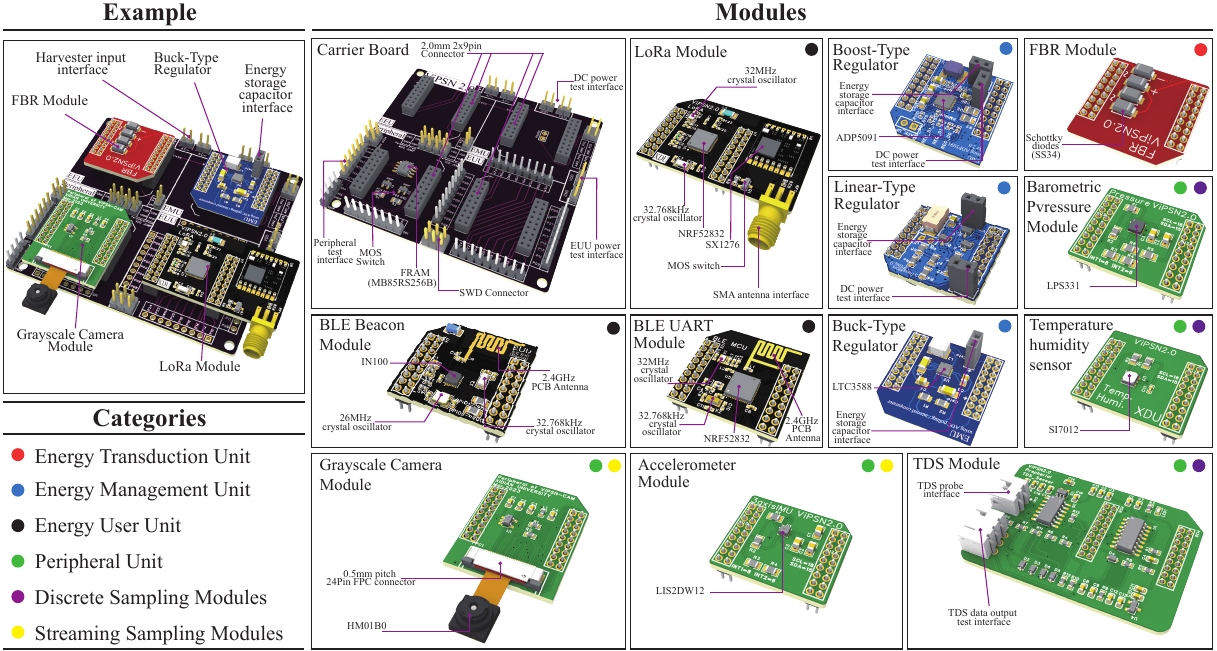}
    \caption{Modular system design and prototype example of ViPSN 2.0. The plug-and-play design enables flexible configuration and rapid deployment for diverse sensing applications. An example assembly demonstrates the integration of different modules on the carrier board, highlighting the system’s versatility and expandability.
    }
    \label{power_compare}
\end{figure*}

\begin{table*}[t]
\centering
\caption{Configurations and Task Power Consumption Statistics of System Components}
\label{table_power_stats}
\renewcommand{\arraystretch}{1.18}
\begin{tabularx}{\textwidth}{p{1.7cm} p{3.7cm} p{3.3cm} X p{2cm} p{2cm}}
\toprule
\textbf{Unit} & \textbf{Module} & \textbf{Parameter} & \textbf{Setting} & \textbf{Task Energy} & \textbf{Static Power} \\
\midrule
\multirow{9}{*}{EUU}
    & \multirow{3}{*}{BLE Beacon Module}
        & Transmission Baudrate & 5400 bps & \multirow{3}{*}{18.9 $\mu$J} & \multirow{3}{*}{2.06 $\mu$W} \\
    &   & Packet Length & 27 Bytes &  &  \\
    &   & Transmission Range & 0--100 m &  &  \\
\cline{2-6}
    & \multirow{4}{*}{BLE UART Module}
        & Transmission Baudrate & 49.4 kbps & \multirow{4}{*}{1.15 mJ} & \multirow{4}{*}{6.6 $\mu$W} \\
    &   & Packet Length & 0--244 Bytes &  &  \\
    &   & Transmission Range & 0--100 m &  &  \\
    &   & Connection Interval & 40 ms &  &  \\
\cline{2-6}
    & \multirow{3}{*}{LoRa Module}
        & Transmission Baudrate & 9600 bps & \multirow{3}{*}{42.84 mJ} & \multirow{3}{*}{6.6 $\mu$W} \\
    &   & Packet Length & 0--100 Bytes &  &  \\
    &   & Transmission Range & 0--2 km &  &  \\
\midrule
\multirow{13}{*}{Peripherals}
    & \multirow{3}{*}{On-chip Temperature Module}
        & Data Length & 32 bits & \multirow{3}{*}{103.68 $\mu$J} & \multirow{3}{*}{6.6 $\mu$W} \\
    &   & Range & -40--85 $^\circ$C &  &  \\
    &   & Accuracy & $\pm$5 $^\circ$C &  &  \\
\cline{2-6}
    & \multirow{6}{*}{Temperature/Humidity Module}
        & \multirow{2}{*}{Data Length} & \parbox[t]{\hsize}{14 bits (Temp.)\\12 bits (Humi.)} & \multirow{6}{*}{673 $\mu$J} & \multirow{6}{*}{0.2 $\mu$W}
        \\
    &   & \multirow{2}{*}{Range} & \parbox[t]{\hsize}{-40--85 $^\circ$C (Temp.)\\0--100 (Humi.)} &  &  \\
    &   & \multirow{2}{*}{Accuracy} & \parbox[t]{\hsize}{$\pm$0.3 $^\circ$C (Temp.)\\$\pm$2\%RH (Humi.)} &  &  \\
\cline{2-6}
    & \multirow{3}{*}{TDS Module}
        & Data Length & 10 bits & \multirow{3}{*}{1.46 mJ} & \multirow{3}{*}{--} \\
    &   & Range & 0--1000 ppm &  &  \\
    &   & Accuracy & $\pm$5\% F.S. &  &  \\
\cline{2-6}
    & \multirow{3}{*}{Barometric Pressure Module}
        & Data Length & 24 bits & \multirow{3}{*}{705.6 $\mu$J} & \multirow{3}{*}{1.65 $\mu$W} \\
    &   & Range & 26--126 kPa &  &  \\
    &   & Accuracy & $\pm$0.02 kPa &  &  \\
\cline{2-6}
    & \multirow{3}{*}{Accelerometer Module}
        & Data Length & 12 bits $\times$ 3 axis & \multirow{3}{*}{608.4 $\mu$J} & \multirow{3}{*}{0.16 $\mu$W} \\
    &   & Range & -2g--2g &  &  \\
    &   & Accuracy & $\pm$20 mg &  &  \\
\cline{2-6}
    & \multirow{2}{*}{Grayscale Camera Module}
        & Image Size & 121$\times$162 & \multirow{2}{*}{16.0 mJ} & \multirow{2}{*}{200 $\mu$W} \\
    &   & Pixel Format & Grayscale 8-bit &  &  \\
\midrule
\multirow{3}{*}{Carrier Board}
    & \multirow{3}{*}{Non-volatile Memory}
        & Type & FRAM & \multirow{3}{*}{\parbox[t]{\hsize}{R: 205.7 $\mu$J\\W: 206.8 $\mu$J}} & \multirow{3}{*}{ 29.7 $\mu$W} \\
    &   & Capacity & 32 KB &  &  \\
    &   & Operation & Single R/W 100 Bytes &  &  \\
\bottomrule
\end{tabularx}
\end{table*}

\section{Modules of ViPSN~2.0}
ViPSN~2.0 adopts a modular hardware architecture, organized around a standardized carrier board and four functional units: the Energy Transduction Unit (ETU), Energy Management Unit (EMU), Energy User Unit (EUU), and a set of interchangeable Peripheral Unit. This design partitioning enables flexible configuration and rapid adaptation to a wide range of vibration-powered IoT scenarios. Each unit is designed with defined electrical and mechanical interfaces, ensuring straightforward integration and reliable operation. This modular approach enhances system scalability and maintainability, providing a robust foundation for diverse deployment scenarios.

\subsection{ETU}
The ETU is responsible for converting the alternating current (AC) generated by VEH transducers into usable direct current (DC) for the EMU. In ViPSN~2.0, a standard AC-to-DC full-bridge rectifier module (FBR) is adopted to achieve robust performance and broad compatibility with various vibration sources encountered in real-world environments. While advanced rectification schemes, such as synchronous electric charge extraction~\cite{Lefeuvre2005-sa} and synchronized switch harvesting on inductor~\cite{Guyomar2005-vb}, can deliver higher efficiency—particularly for PZT transducers under resonant excitation—their effectiveness tends to diminish under variable or transient vibration conditions. Thus, the conventional rectification approach is chosen as a practical and reliable baseline, ensuring consistent energy transduction across multiple real-world application scenarios.

\subsection{EMU}
The EMU in ViPSN~2.0 provides efficient DC-DC regulation tailored to the requirements of different VEH sources. Three interchangeable regulator modules—buck, linear, and boost—enable rapid adaptation to varying transducer output profiles and application scenarios. Developers can select and configure the most suitable module to optimize energy conversion for their specific deployment.

\subsubsection{Buck-Type Regulator}
The buck-type regulator (LTC3588) is designed for energy harvesting sources with relatively high output voltages (2.7~V-20~V). It employs burst-mode operation to maintain high efficiency under light-load conditions and integrates a low-quiescent-current rectifier with adjustable output voltage. The built-in under-voltage lockout protection ensures stable operation during input voltage fluctuations. The output voltage can be configured via pin headers, allowing adaptation to different system requirements.

\subsubsection{Linear-Type Regulator}
The linear-type regulator utilizes a depletion-mode MOSFET (BSS159N) pass element to maintain stable, low-noise output voltage without high-frequency switching. This simple architecture results in low electromagnetic interference and fast transient response, which is particularly beneficial for noise-sensitive, battery-free IoT applications. Although efficiency decreases with large input-to-output voltage differences, the robustness and low-noise performance make linear regulation effective for low-power scenarios where voltage stability is critical.

\subsubsection{Boost-Type Regulator}
The boost-type regulator (ADP5091) is designed for low-voltage energy sources (80~mV-3.6~V). It integrates maximum power point tracking (MPPT) to improve extraction efficiency under intermittent and variable input conditions, and features cold-start circuitry that enables operation from very low input voltages. Both the output voltage and MPPT parameters can be configured via external resistors, providing flexibility to accommodate a wide range of energy harvesting environments.

\subsection{EUU}
The EUU offers wireless communication modules tailored to various data throughput requirements and transmission ranges, ensuring reliable and efficient operation under ambient energy availability.

\subsubsection{BLE Beacon Module}
The ultra-low-power Bluetooth Low Energy (BLE) beacon module is optimized for energy-constrained, battery-free systems. The energy required for each transmission can be as low as approximately 19~$\mu$J (Table~\ref{table_power_stats}), supporting periodic or event-driven broadcasting of concise data such as temperature, humidity, or device identification via 27-byte beacon packets. Its lightweight protocol and minimal activation overhead enable reliable operation even under highly intermittent power conditions. The unified interface allows developers to deploy and replace the module without additional configuration.

\subsubsection{BLE UART Module}
For applications that require higher data throughput, the BLE UART module provides a flexible solution. Built on the nRF52832 system-on-chip (SoC), it achieves static power consumption as low as 6.6~$\mu$W (Table~\ref{table_power_stats}) and supports programmable, timer-based wake-up. This design enables efficient scheduling and power management during intermittent operation, while the standardized module interface simplifies both system integration and firmware development.

\subsubsection{LoRa Module}
The Long Range (LoRa) module, based on the SX1276 transceiver and paired with an nRF52832 SoC as the microcontroller unit (MCU), supports wireless communication over multi-kilometer distances. While each transmission consumes more energy than BLE, the extended range minimizes the need for multi-hop relays and infrastructure, making it ideal for remote or spatially distributed, battery-free applications. The platform automatically detects and configures the installed radio module, allowing developers to seamlessly switch between short-range and long-range communication with minimal hardware effort.

\subsection{Peripherals}
ViPSN~2.0 provides standardized pluggable interfaces for both discrete and streaming sensors, enabling rapid prototyping and flexible adaptation to diverse sensing needs.

\subsubsection{Discrete Sampling Modules}
Discrete sampling modules target low-power, periodic data collection scenarios, including on-chip temperature module (built on the nRF52832 SoC), temperature/humidity module (SI7021), barometric pressure module (LPS331), and total dissolved solids (TDS) water quality module. All modules adopt a pluggable header interface, allowing developers to easily insert, replace, or upgrade sensing units without hardware modifications. These low-power sensors are well-suited to the constraints of vibration-powered operation. Detailed energy consumption profiles and sensor configuration parameters are summarized in Table~\ref{table_power_stats}.

\subsubsection{Streaming Sampling Modules}
To support continuous or high-frequency sensing, ViPSN~2.0 integrates streaming-capable peripherals for inertial and visual data acquisition. The onboard 3-axis accelerometer module (LIS2DW12) enables efficient collection of vibration signals. Additionally, the low-power grayscale camera module (HM01B0) supports image capture at a resolution of 121×162 pixels (approximately 20,000 pixels), enabling basic visual recognition tasks within strict energy constraints. These streaming sensors are suitable for battery-free operation, and their measured energy profiles (Table~\ref{table_power_stats}) demonstrate the feasibility of incorporating real-time data sources into vibration-powered IoT systems.

\subsection{Carrier Board}
The carrier board forms the physical and electrical backbone of ViPSN~2.0, interconnecting all functional modules through standardized, pluggable interfaces. In addition to reliable interconnection, it integrates essential support features for robust intermittent operation, including non-volatile memory and multiple expansion and test ports. Mechanical alignment notches and keyed pin headers ensure correct assembly and enhance system robustness during prototyping and deployment. 

\subsubsection{Non-Volatile Memory}
The carrier board integrates a 32~kB ferroelectric random access memory (FRAM, MB85RS256B) to provide low-power, non-volatile data storage, ensuring reliable retention of sensor data and system states across intermittent power cycles. Owing to its high write speed and virtually unlimited endurance, FRAM is well-suited for battery-free IoT applications. The energy required to read or write 100 bytes of data is summarized in Table~\ref{table_power_stats}.

\subsubsection{Peripheral Ports}
The carrier board provides multiple standardized sensor interfaces (I²C, UART, SPI), enabling flexible expansion tailored to specific application requirements. The modular and open interface facilitates easy integration of additional peripherals and communication modules beyond provided examples.

\subsubsection{Mechanical Design}
The board employs four 2×9 pin headers (2.0~mm pitch) to connect all functional modules. Alignment notches and matching PCB cutouts minimize assembly errors and ensure reliable integration, supporting both field deployment and future module upgrades.

\section{Power Solutions of ViPSN~2.0}
\label{Power Solutions}
Effective power management is crucial for vibration-powered IoT systems to ensure robust and reliable operation under intermittent and unpredictable vibration conditions. It provides temporary capacitive energy storage for harvested vibration energy and delivers stable, regulated voltage to system loads and sensors. As vibration sources typically exhibit intermittent, transient, and variable characteristics, the power management circuit must efficiently maximize harvested energy, minimize conversion losses, and reliably deliver power according to dynamic load demands. Additionally, effective checkpointing mechanisms under intermittent execution require explicit energy-level awareness—a capability insufficiently supported by conventional power management solutions.

This section presents three representative solutions—under-voltage lockout (UVLO), passive interrupt detection (PID), and active polling check (APC)—that address these challenges and provide comprehensive power solutions tailored specifically for vibration-powered IoT applications.

\begin{figure}[t]
\includegraphics[width=\columnwidth]{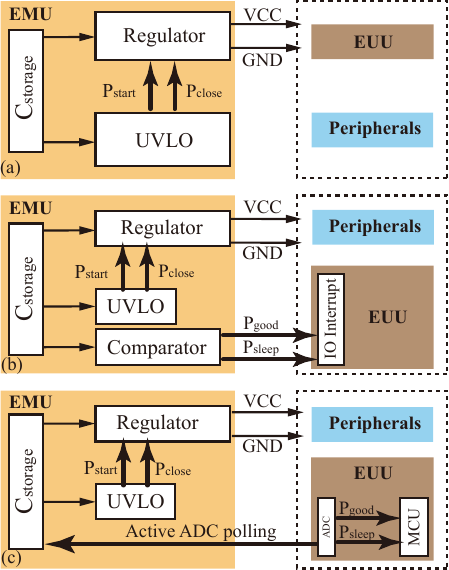}\centering
\caption{Schematic of power solutions for vibration-powered iot systems: (a) UVLO solution with regulator; (b) PID solution with interrupt detection; (c) APC solution with active polling check.}
\label{power_compare}
\end{figure}

\subsection{UVLO Solution}
The first solution is based on the UVLO mechanism, a fundamental power management strategy that is widely implemented in linear-, boost-, and buck-type regulators, as illustrated in Fig.~\ref{power_compare}. The UVLO mechanism provides two inherent regulator signals: $\mathrm{P_{start}}$ and $\mathrm{P_{close}}$. However, these fixed UVLO thresholds ($V_\text{Pstart}$ and $V_\text{Pclose}$) are determined solely by the regulated output voltage conditions, and therefore do not explicitly reflect the actual stored energy level in the storage capacitor. As a result, the system’s energy state can only be divided into two phases: the cold-start phase ($0\,\mathrm{V}$ to $V_\text{Pstart}$) and the task-operation phase ($V_\text{Pstart}$ to $V_\text{Pclose}$).

Due to this limitation, selecting an appropriate storage capacitor becomes critical to achieving a balanced trade-off among storage capacity, charging responsiveness, and operational reliability. The usable energy available is determined by these fixed UVLO thresholds, and is given by $\frac{1}{2}C_\text{storage}(V_\text{Pstart}^2-V_\text{Pclose}^2)$, where $V_\text{Pstart}$ corresponds to the regulator turn-on event, and $V_\text{Pclose}$ indicates the regulator shutdown event.

To ensure sufficient energy for system initialization and execution of energy-intensive tasks, the chosen storage capacitor must satisfy the following condition:
\begin{eqnarray}
\label{e1}
C_\text{storage} \ge \frac{2\left[\int_{T_i}{P_\text{static}(t)\text{d}t}+E_{\text{task},i}\right]}{\eta \left(V_{\text{Pstart}}^2 - V_\text{Pclose}^2\right)},
\end{eqnarray}
where $C_\text{storage}$ is the storage capacitance, $P_{\text{static}}(t)$ is the static power consumption of the system, $\eta$ represents the average DC-DC conversion efficiency, and $E_{\text{task},i}$ denotes the total energy required by critical tasks including initialization, sensing and transmitting.

\subsection{PID Solution}
The PID solution is an enhanced power management circuit that provides explicit energy-level signals, as adopted by many applications~\cite{hester2017flicker,li2020vipsn}. 

As shown in Fig.~\ref{power_compare}(b), a comparator continuously monitors the voltage of the storage capacitor. Two additional energy-level signals, $\mathrm{P_{good}}$ and $\mathrm{P_{sleep}}$, are introduced beyond the regulator's original signals ($\mathrm{P_{start}}$ and $\mathrm{P_{close}}$). Specifically, signal $\mathrm{P_{good}}$ is passively generated by the hardware at the rising edge of the comparator's output, indicating that the storage capacitor has accumulated sufficient energy to execute the most power-consuming atomic operation. Conversely, signal $\mathrm{P_{sleep}}$ is generated at the falling edge of the comparator's output, serving as a passive interrupt to the EUU and warning of imminent energy shortage. Upon receiving $\mathrm{P_{sleep}}$, the EUU is promptly interrupted and must initiate an emergency procedure to save critical data into non-volatile memory and subsequently enter an ultra-low-power deep-sleep mode.

Moreover, developers can adjust the voltage thresholds for generating $\mathrm{P_{good}}$ and $\mathrm{P_{sleep}}$ by configuring the comparator's resistor network, ensuring proper operation under various excitation conditions. With these adjustable thresholds, the available energy is explicitly divided into the task-operation phase ($V_\text{Pgood}$ to $V_\text{Psleep}$), i.e., $\frac{1}{2}C_\text{storage}(V_\text{Pgood}^2 - V_\text{Psleep}^2)$, which provides sufficient energy for normal operations such as restoring from checkpoints, sensing, and wireless communication; and the checkpoint phase ($V_\text{Psleep}$ to $V_\text{Pclose}$), i.e., $\frac{1}{2}C_\text{storage}(V_\text{Psleep}^2 - V_\text{Pclose}^2)$, which reserves energy specifically for checkpointing operations, namely saving the program’s address space, stack, registers, and global variables into non-volatile memory, thereby ensuring a consistent state-saved starting point upon energy restoration.

\subsection{APC Solution}
The APC solution is based on periodic on-chip ADC sampling, as illustrated in Fig.~\ref{power_compare}(c). This approach is widely adopted in intermittent execution software systems because it can be directly implemented on existing hardware platforms without requiring additional hardware modifications~\cite{ransford2011mementos,balsamo2014hibernus,jayakumar2014quickrecall}.

In this solution, the system periodically samples the storage capacitor voltage using the ADC and compares it with a predefined threshold. As long as the measured voltage remains above the threshold, normal operations proceed without checkpointing. Once the measured voltage falls below the threshold, imminent power failure is detected and checkpointing procedures are immediately triggered.

However, for power-limited VEH systems, frequent ADC polling introduces considerable overhead in energy consumption and execution time. To ensure timely detection of energy depletion, adequate sampling points must be inserted into the program at compile time, facilitating effective runtime monitoring of energy trends. This issue is particularly severe under intermittent and unpredictable vibration conditions, where the measurement overhead itself can exceed the harvested energy, potentially hindering the completion of critical tasks.

In this solution, the system's static power consumption $P_{\text{static}}(t)$ is defined as:
\begin{eqnarray}
\label{e2}
P_\text{static}(t)=f_s E_\text{adc} + P_\text{dc-dc}(t) + P_\text{leak}(t),
\end{eqnarray}
where $f_s$ denotes the ADC polling frequency, determined by desired energy measurement accuracy and vibration conditions;
$E_{\text{adc}}$ is the energy consumption per ADC sampling;
$P_{\text{dc-dc}}(t)$ denotes the regulator's operating power;
and $P_{\text{leak}}(t)$ represents the leakage power caused by the storage capacitor's equivalent parallel leakage resistance.

\begin{figure}[htbp]
\includegraphics[width=\columnwidth]{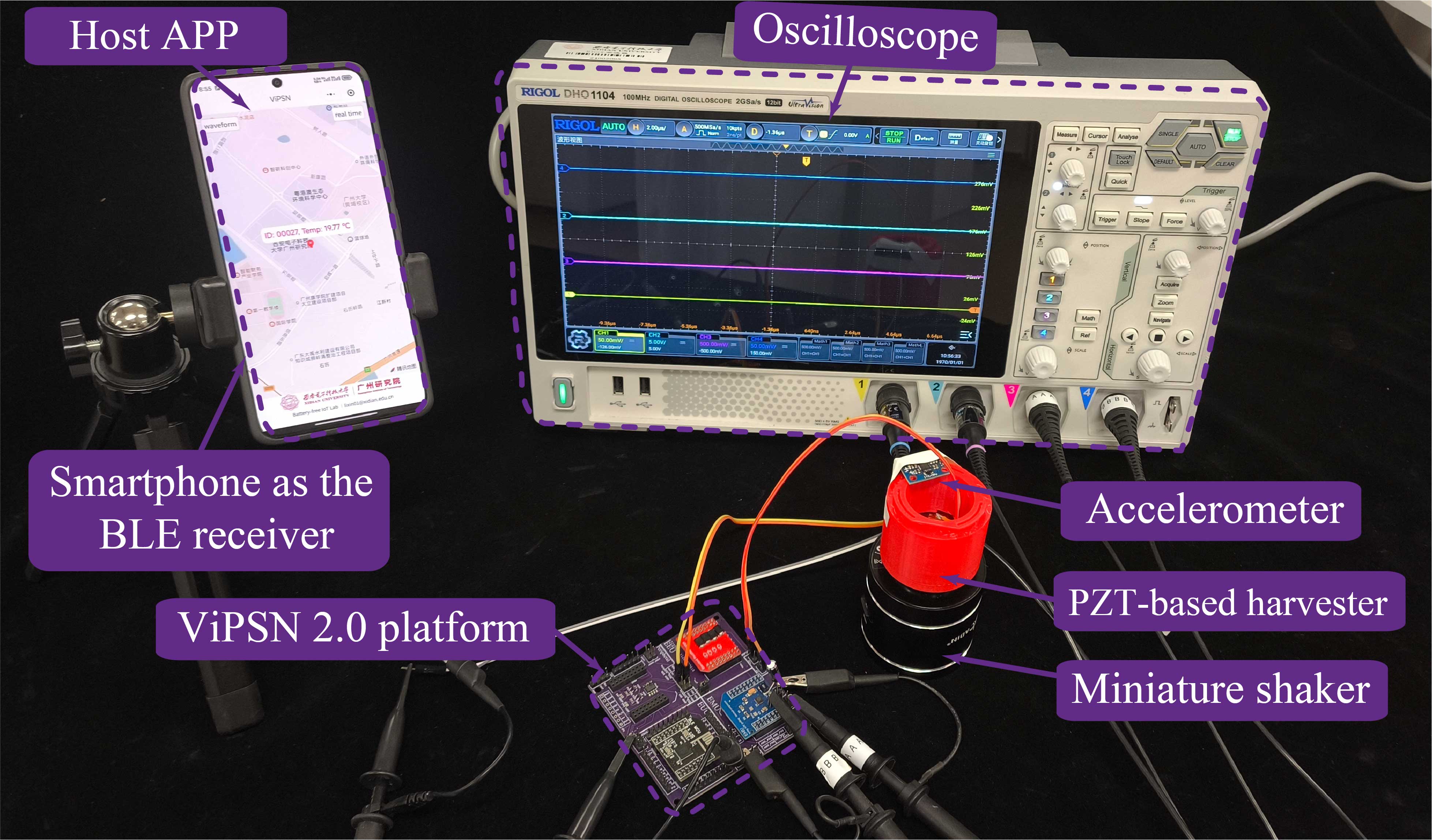}\centering
\caption{Photograph of the experimental setup. The smart phone acts as a BLE receiver host to collect beacon packets transmitted from the ViPSN~2.0 prototype. The shaker is used to simulate the vibration excitation of the Clifton Suspension Bridge.}
\label{prototype}
\end{figure}

\begin{figure*}[!ht]
\includegraphics[width=2\columnwidth]{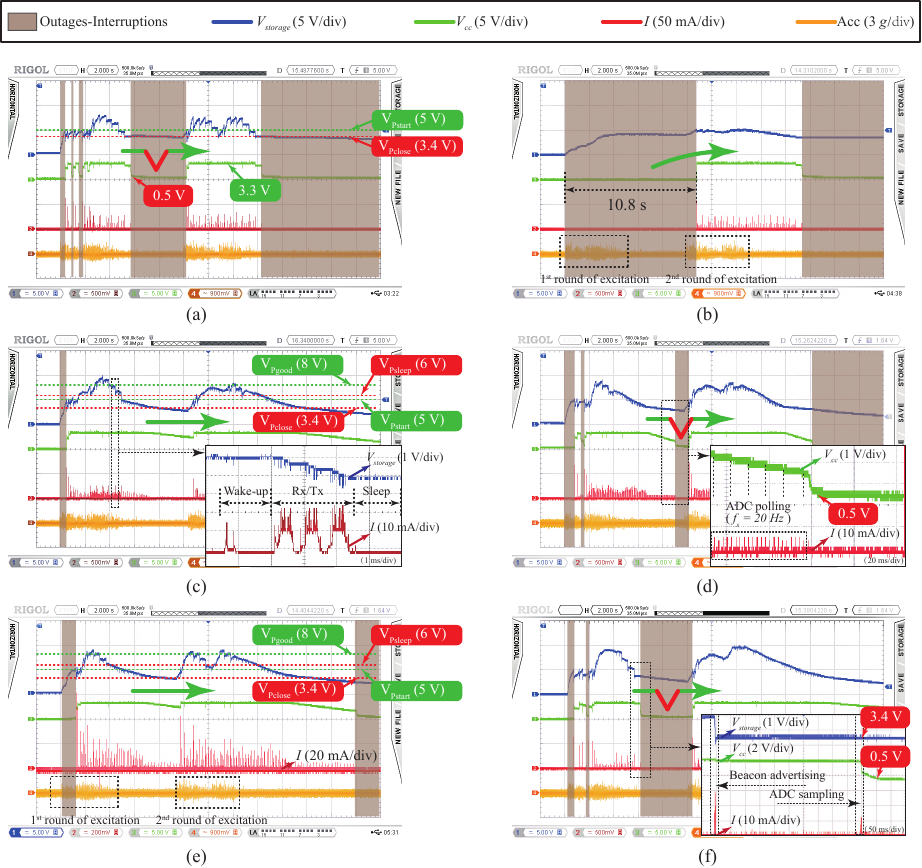}\centering
\caption{Waveforms of the vibration-powered temperature sensor prototype with three different power solutions: UVLO (a, b), PID (c), and APC (d--f).
(a) Startup with outage using a small storage (10 $\mu$F).
(b) Success but slow startup without outage using a large storage (100 $\mu$F).
(c) Execution without outage using energy indicating interrupt signals.
(d) Startup with outage due to over-sampling ($f_s$ = 20 Hz).
(e) Execution without outage using ADC polling check ($f_s$ = 4 Hz).
(f) Execution with outage due to under-sampling ($f_s$ = 0.5 Hz).
The frequency of the BLE Beacon advertisement is 2 Hz.
(c) - (f) use a small storage capacitor (10 $\mu$F).
$V_{\text{Pstart}}$ and $V_{\text{Pclose}}$ are fixed thresholds, while $V_\text{Pgood}$ and $V_\text{Psleep}$ are tunable according to the most energy-consuming atomic operation.}
\label{waveform}
\end{figure*}

\subsection{Experimental Validation}
In this subsection, we experimentally validate the three proposed power solutions—UVLO, PID, and APC—implemented in ViPSN~2.0. The aim is to clearly demonstrate the performance differences and suitable application scenarios for each solution under realistic vibration conditions.

\subsubsection{Setup}
Fig.~\ref{prototype} illustrates the hardware setup used for experimental validation. Using modules from ViPSN~2.0, we assembled a vibration-powered temperature sensor, its modular setup is as follows:
\begin{itemize}
    \item ETU: PZT (35~mm diameter piezoelectric disk with a 10~g proof mass) + FBR;
    \item EMU: Buck-Type Regulator, with a 10~$\mu$F storage capacitor;
    \item EUU: BLE Beacon Module (advertisement interval: 500~ms);
    \item Peripheral: On-chip Temperature Module, serving as a representative sensing load.
\end{itemize}
Real-world vibration conditions are reproduced using vibration data obtained from the Energy Harvesting Network Data Repository\footnote{\url{http://eh-network.org/data/}}. Specifically, we select a 24-second vibration segment recorded near the pillars of the Clifton Suspension Bridge, clearly showing two separate excitation events caused by passing vehicles, as illustrated in Fig.~\ref{waveform}.

\subsubsection{Configuration}
To systematically evaluate the three power solutions, the EMU module is configured as follows:
\begin{itemize}
    \item UVLO: Configured to use only the built-in UVLO mechanism of the buck-type regulator, which features fixed internal voltage thresholds ($V_{\text{Pstart}}=5~\text{V}$, $V_{\text{Pclose}}=3.6~\text{V}$) that directly determine system startup and shutdown events.
    \item PID: Configured with an external micropower comparator (MIC841) connected to the regulator. The comparator generates explicit energy-level signals ($\mathrm{P_{good}}$ and $\mathrm{P_{sleep}}$) based on adjustable voltage thresholds, providing clear indication of whether the stored energy is sufficient. These signals enable reliable checkpointing and subsequent deep-sleep transitions.
    \item APC: Configured to employ the integrated ADC of the BLE Beacon module for periodic measurement of the storage capacitor voltage. Software-based polling dynamically monitors energy levels against predefined thresholds ($V_{\text{Pgood}}$ and $V_{\text{Psleep}}$), determining the optimal timing for checkpointing and deep-sleep operations.
\end{itemize}
These configurations enable a direct and fair comparison of the power solutions under identical excitation conditions.

\begin{table}[htbp]
\centering
\caption{Power Consumption of Different Power Solutions}
\label{table:energy_comparison}
\renewcommand{\arraystretch}{1.16}
\begin{tabularx}{\linewidth}{
    >{\raggedright\arraybackslash}m{1.7cm}
    >{\raggedright\arraybackslash}X
    >{\centering\arraybackslash}m{3.3cm}
}
\toprule
\textbf{Solution} & \textbf{Operation} & \textbf{Power Consumption} \\
\midrule

\multirow{2}{*}{UVLO}
    & Initialization          & 50.8~$\mu$J \\
    & Static Power            & 5.0~$\mu$W  \\
\midrule

\multirow{2}{*}{PID}
    & Initialization          & 50.9~$\mu$J \\
    & Static Power            & 28.0~$\mu$W \\
\midrule

\multirow{5}{*}{APC}
    & Initialization         & 51.09~$\mu$J \\
    & ADC Sampling            & 1.29~$\mu$J \\
    & Polling @ 0.5~Hz  & 27.65~$\mu$W \\
    & Polling @ 4~Hz    & 32.16~$\mu$W \\
    & Polling @ 20~Hz   & 52.80~$\mu$W \\
\bottomrule
\end{tabularx}
\end{table}

\subsubsection{Power Consumption Analysis}
The power and energy statistics for different tasks are summarized in Table~\ref{table:energy_comparison}. Before performing temperature sampling and wireless transmission tasks, the circuits of the EMU, EUU, and peripheral modules must undergo an initialization process. Thus, the harvested energy must first satisfy the energy requirements associated with this initialization before task execution can proceed. Table~\ref{table:energy_comparison} lists the energy consumption of different functions under the three power solutions. Each entry reflects the combined contributions of the relevant circuit hardware and software components, including the basic voltage regulator, the energy indication function implemented by an analog comparator (PID), and ADC-based sampling (APC).

For the PID solution, static power consumption includes both the comparator and the regulator. For the APC solution, the static power is mainly determined by the ADC sampling frequency, as summarized in Eq.~(\ref{e2}). The UVLO solution has the lowest static power overhead, as it does not require additional monitoring circuits.

To guarantee task completion without energy outage, it is necessary to accumulate enough energy in the storage capacitor and execute computational tasks in bursts, rather than relying solely on the input power. According to Eq.~(\ref{e1}) and the task requirements, the minimum storage capacitances for the three power solutions are calculated as 6.76~$\mu$F (UVLO), 7.59~$\mu$F (PID), and 7.74~$\mu$F (APC), respectively. Therefore, in this study, a 10~$\mu$F capacitor is used as the storage device for all solutions.

\subsubsection{Intermittent Execution}
We evaluate the number of outages—interruptions in workload execution due to insufficient stored energy—for the three power solutions implemented on ViPSN~2.0. Outage frequency is a key index of efficient program execution: the more outages, the more energy is spent on checkpointing and recovery rather than useful computations. Reducing unnecessary checkpointing before and after outages enables more efficient application progress under the same vibration excitation.

Fig.~\ref{waveform} shows the operational waveforms for the three power solutions. For the UVLO solution, with only a basic regulator, outages are unpredictable, and there is no energy reference for controlling code execution. Without explicit energy indication, checkpointing operations—such as stopping energy-consuming functions or entering deep-sleep mode—cannot be performed in time to save energy. The available energy is determined by the fixed $V_\text{Pstart}$ and $V_\text{Pclose}$ thresholds. With a small storage capacitance (e.g., 10~$\mu$F), reliable startup is not possible. Using a larger capacitance (e.g., 100~$\mu$F) can avoid outages (red area in Fig.~\ref{waveform}(b)), but results in a much longer cold-start period, reaching up to 10.8~s.

By adding energy-indicating signals, checkpoint-capable solutions (PID and APC) increase the usable energy for code execution from $\frac{1}{2}C_\text{storage}(V_\text{Pstart}^2 - V_\text{Pclose}^2)$ to $\frac{1}{2}C_\text{storage}(V_\text{Pgood}^2 - V_\text{Pclose}^2)$. This ensures sufficient stored energy for reliable startup. Additionally, by detecting when the storage voltage is between $V_\text{Psleep}$ and $V_\text{Pclose}$, necessary checkpointing can be performed and the system robustly switches between computation and deep-sleep energy build-up phases. During energy build-up phase, the system remains in hibernation with basic awareness, enabling prompt wake-up on the next excitation. The successful startup process with energy signals is shown in Fig.~\ref{waveform}(c).

For active polling, periodic comparison of $V_\text{storage}$ with software-defined thresholds provides $\mathrm{P_{good}}$ and $\mathrm{P_{sleep}}$ energy signals. When the rising edge of $V_\text{storage}$ crosses $V_\text{Pgood}$, beacon packets are broadcast. When the falling edge crosses $V_\text{Psleep}$, imminent power failure is indicated and checkpointing is triggered. Fig.~\ref{waveform}(e) shows successful execution without outages between two periods of intensive excitation.

\subsubsection{Performance and Overhead}
Compared to the PID solution, timely awareness of stored energy in the APC scheme depends on the ADC sampling frequency. Fig.~\ref{waveform}(d)-(f) shows system performance under three polling frequencies: 20~Hz, 4~Hz, and 0.5~Hz. For the 20~Hz case, the polling energy overhead exceeds the energy harvested from the vibration source in the same period, resulting in an over-sampling scenario. As static power consumption ($P_\text{static}$) increases, even if the energy-warning signal (${P_\text{sleep}}$) is detected, the remaining stored energy $\frac{1}{2}C_\text{storage}(V_\text{Psleep}^2 - V_\text{Pclose}^2)$ is often insufficient to maintain deep-sleep until the next excitation, causing the system to shut down unexpectedly for a short period, as shown in Fig.~\ref{waveform}(d).

On the other hand, when the sampling rate is low (0.5~Hz), the system may miss the critical moment to perform checkpointing if $V_\text{storage}$ drops rapidly. This under-sampling case can also result in missed checkpoints and subsequent outages, as shown in Fig.~\ref{waveform}(f). Thus, the ADC polling frequency is highly sensitive for maintaining robust and reliable operation. A moderate sampling rate (4~Hz) provides a good balance, minimizing both power overhead and the risk of missed checkpoints, ensuring stable intermittent execution without outages (Fig.~\ref{waveform}(e)).

\subsubsection{Power Solution Selection Guidelines}
Although sensitive to polling frequency, the APC solution offers unique flexibility through software-defined checkpoint thresholds, enabled by high-resolution ADC sampling. This feature allows for in-situ adjustment of both thresholds and sampling rates, making APC more adaptable to complex-duty tasks, such as image or audio streaming, computation-intensive workloads, or frequent wireless transmissions. 

In contrast, the PID solution is more appropriate for relatively stable heavy-duty tasks, where ultra-low-power operation is essential and frequent ADC measurements are undesirable. 

The UVLO solution, characterized by its simplicity and zero additional overhead, is most suitable for light-duty tasks.

\begin{figure}[htbp]
    \centering
    \includegraphics[width=\columnwidth]{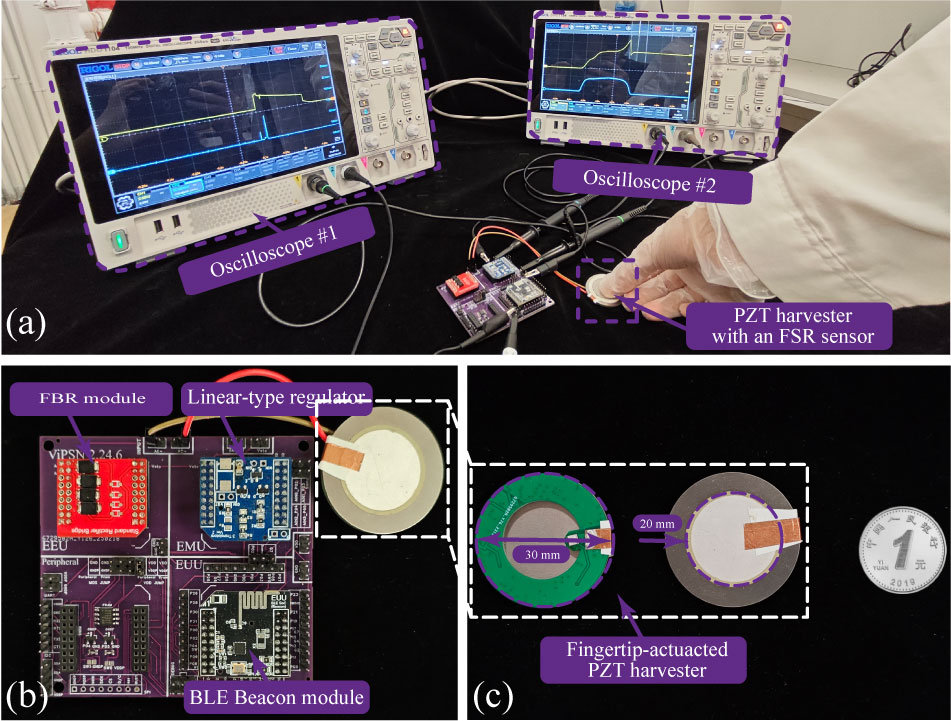}
    \caption{
    Prototype and experimental setup for the ViPSN-Beacon case.
    (a) ViPSN-Beacon test setup with an FSR sensor configured to monitor the pressure applied by a single finger on the PZT harvester.
    (b) Prototype device.
    (c) PZT harvester dimensions ($\phi$35 mm with a piezoelectric ceramic disk of $\phi$20 mm, cost less than \$0.002).
    }
    \label{fig:fig_case_beacon_pzt}
\end{figure}

\begin{table}[htbp]
\small
\centering
\caption{System Configuration and Power Consumption of ViPSN-Beacon}
\label{tab:system_camera_results}
\begin{tabularx}{\linewidth}{p{4.2cm} X}
\hline
\textbf{Parameter} & \textbf{Value} \\
\hline
Deployment location & Indoor office and outdoor campus \\
\textit{Excitation source} & \textit{Single fingertip actuation cycle} \\
Harvester transducer & PZT \\
Average harvester output power & $125~\mu$W\\
Harvester open-circuit voltage & $\sim 110$ V \\
Harvester short-circuit current & $\sim 50~\mu$A \\
\hline
Storage capacitance & $2.2~\mu$F \\
\hline
Transmission protocol & BLE Beacon, 2.4~GHz \\
Transmission power & 0~dbm \\
Transmission data payload & 27~Bytes \\
\textit{Transmission range} & \textit{100~m} \\
\hline
Initialization energy & 9.017~$\mu$J \\
Transmission energy & 9.860~$\mu$J \\
\hline
Total energy consumption & 18.9~$\mu$J \\
\hline
\end{tabularx}
\end{table}

\begin{figure*}[t]
    \centering
    \includegraphics[width=2\columnwidth]{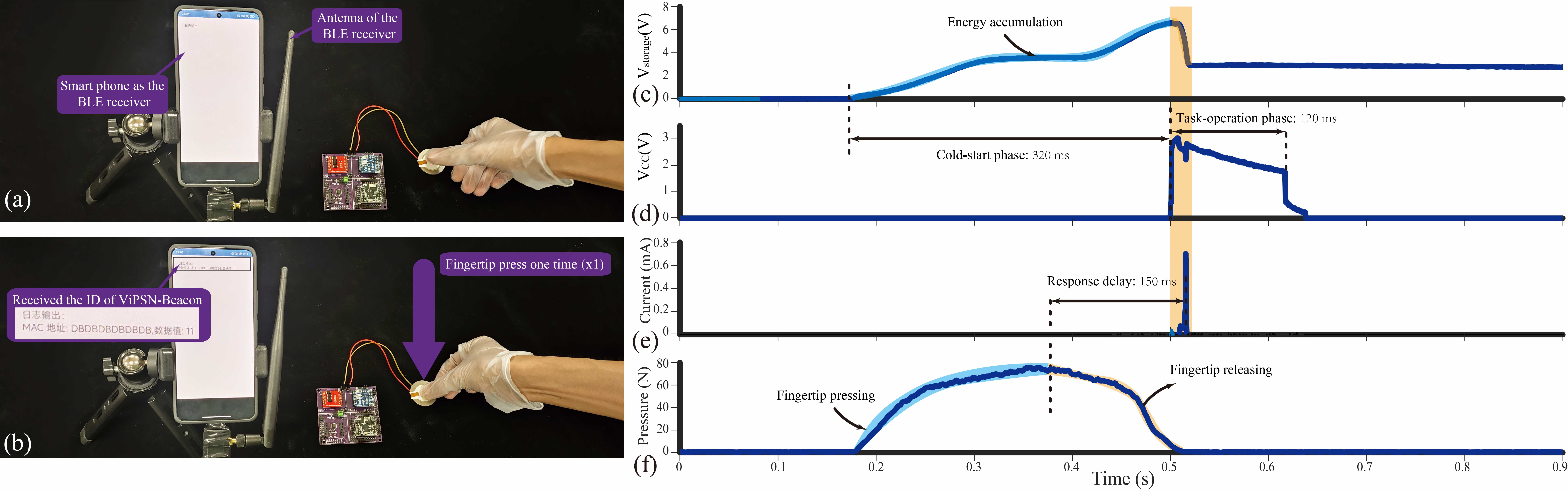}
    \caption{
        System response evaluation of the ViPSN-Beacon. 
        (a) Preparation for fingertip actuation with BLE receiver on smartphone.
        (b) Fingertip actuation with successful ViPSN-Beacon ID reception on smartphone.
        (c)--(f) Measured parameters during a single actuation cycle: (d) Storage capacitor voltage $V_\text{storage}$; (e) BLE Beacon module supply voltage $V_{\mathrm{cc}}$; (f) Instantaneous current consumption; (g) Applied force on the PZT harvester.
    }
    \label{fig:fig_case_beacon_test_indoor}
\end{figure*}

\begin{figure*}[htbp]
    \centering
    \includegraphics[width=2\columnwidth]{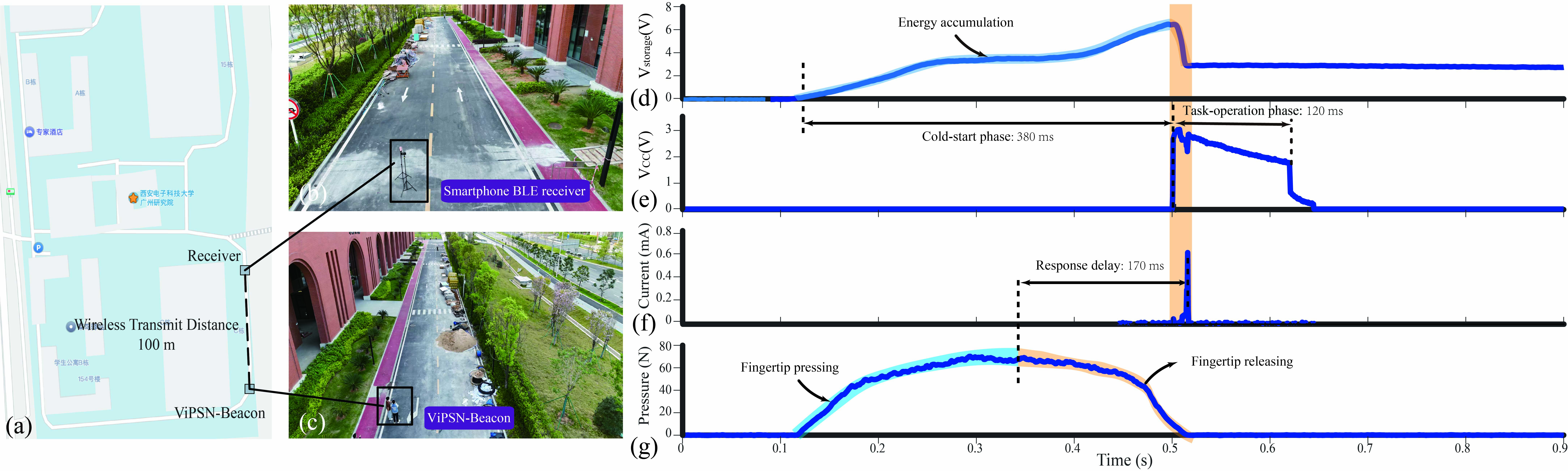}
    \caption{
        Outdoor wireless transmission and system response of the ViPSN-Beacon prototype.
        (a) Map of the experimental area showing the 100~m line-of-sight distance between the ViPSN-Beacon and the BLE receiver.
        (b) Photograph of the smartphone-based BLE receiver setup.
        (c) Photograph of the ViPSN-Beacon (transmitter).
        (d)--(g) Measured parameters during a single actuation cycle: (d) Storage capacitor voltage $V_\text{storage}$; (e) BLE Beacon module supply voltage $V_{\mathrm{cc}}$; (f) Instantaneous current consumption; (g) Applied force on the PZT harvester.
    }
    \label{fig:fig_case_beacon_test_outdoor}
\end{figure*}

\section{ViPSN-Beacon: Fingertip-powered PZT Beacon}
To verify the performance of the UVLO power solution and assess the adaptability of the ViPSN~2.0 platform under highly transient energy harvesting conditions, we developed the ViPSN-Beacon prototype. This prototype operates with a minimalist energy path, harvesting energy from a single fingertip press applied to a PZT transducer. Experimental results indicate that, despite the limited energy and short pulse duration provided by the PZT disc, the system can reliably transmit a BLE Beacon, with a measured maximum detectable range of approximately 100 meters.

\subsection{Setup}
As shown in Fig.~\ref{fig:fig_case_beacon_pzt}, the ViPSN-Beacon consists of three modules:
\begin{itemize}
    \item ETU: PZT (35~mm diameter) + FBR;
    \item EMU: Linear-Type Regulator with the UVLO power solution;
    \item EUU: BLE Beacon Module.
\end{itemize}

The ETU utilizes a low-cost commercial PZT-5 ceramic buzzer disc (less than \$0.002 per unit), with dimensions comparable to a coin. A minimal supporting frame, shown in Fig.~\ref{fig:fig_case_beacon_pzt}(b), ensures reliable assembly and provides adequate compression space for fingertip activation. This simple mechanical structure guarantees consistent and repeatable energy harvesting during operation. To capture the limited energy generated by each pressing event, we employed a linear-type regulator combined with an ultra-low-power UVLO circuit. A 2.2~$\mu$F capacitor buffers the transient high-voltage output from the PZT transducer, balancing rapid energy capture with sufficient charge delivery for wireless transmission. The UVLO thresholds ($V_{\text{Pstart}}=6.7~\text{V}$, $V_{\text{Pclose}}=2.8~\text{V}$) are optimized for the high-voltage, short-duration pulses characteristic of PZT harvesters, ensuring efficient utilization of these brief mechanical events. The EUU consists of a BLE Beacon module configured to transmit identification signals. The adopted power management strategy ensures stable energy delivery, enabling reliable beacon operation under highly transient and low-energy harvesting conditions.

\subsection{Evaluation}
To evaluate the performance of the ViPSN-Beacon under realistic transient energy harvesting conditions, we conducted indoor and outdoor experiments (Fig. \ref{fig:fig_case_beacon_test_indoor}, Fig. \ref{fig:fig_case_beacon_test_outdoor}). In each experiment, a single fingertip actuation cycle was performed, during which we simultaneously measured the EMU storage capacitor voltage ($V_\text{storage}$), EUU supply voltage ($V_{\mathrm{cc}}$), instantaneous EUU current, and the applied fingertip force using a force-sensitive resistor (FSR402).

Fig. \ref{fig:fig_case_beacon_test_indoor}(c)–(f) presents the indoor experimental results. During each actuation cycle—consisting of pressing and releasing actions—the applied force remained below 80 N, which is within the range of force that can be comfortably applied by the fingertip of an average adult. The system's operation is governed by the UVLO-defined thresholds. When $V_\text{storage}$ surpasses the startup threshold $V_{\text{Pstart}} = 6.7~\text{V}$, the system transitions from the cold-start phase into the active task-operation phase. During the approximately 320~ms cold-start phase, energy accumulates within the storage capacitor. Upon activation, the BLE Beacon module powers on, briefly enters active mode, and transmits a beacon packet. The task-operation phase lasts approximately 120~ms, during which a sharp peak in the current trace (Fig.~\ref{fig:fig_case_beacon_test_indoor}(e)) indicates active transmission. Subsequently, the current quickly drops as the BLE Beacon module transitions into an ultra-low-power standby mode with a static power consumption of approximately 2.06~$\mu$W. The task-operation phase ends when $V_\text{storage}$ falls below the shutdown threshold $V_{\text{Pclose}} = 2.8~\text{V}$, returning the system to the shutdown phase.

Outdoor experiments were performed at a line-of-sight distance of 100~m between the ViPSN-Beacon device and a smartphone-based BLE receiver (Fig. \ref{fig:fig_case_beacon_test_outdoor}(a)–(c)). The measured results (Fig. \ref{fig:fig_case_beacon_test_outdoor}(d)–(g)) confirmed successful beacon reception at this distance. The outdoor response latency was approximately 170~ms. These experimental results validate ViPSN-Beacon's robust operation and reliability under practical outdoor conditions, demonstrating its significant potential for real-world transient-powered IoT deployments.

\begin{figure}[t]
    \centering
    \includegraphics[width=\columnwidth]{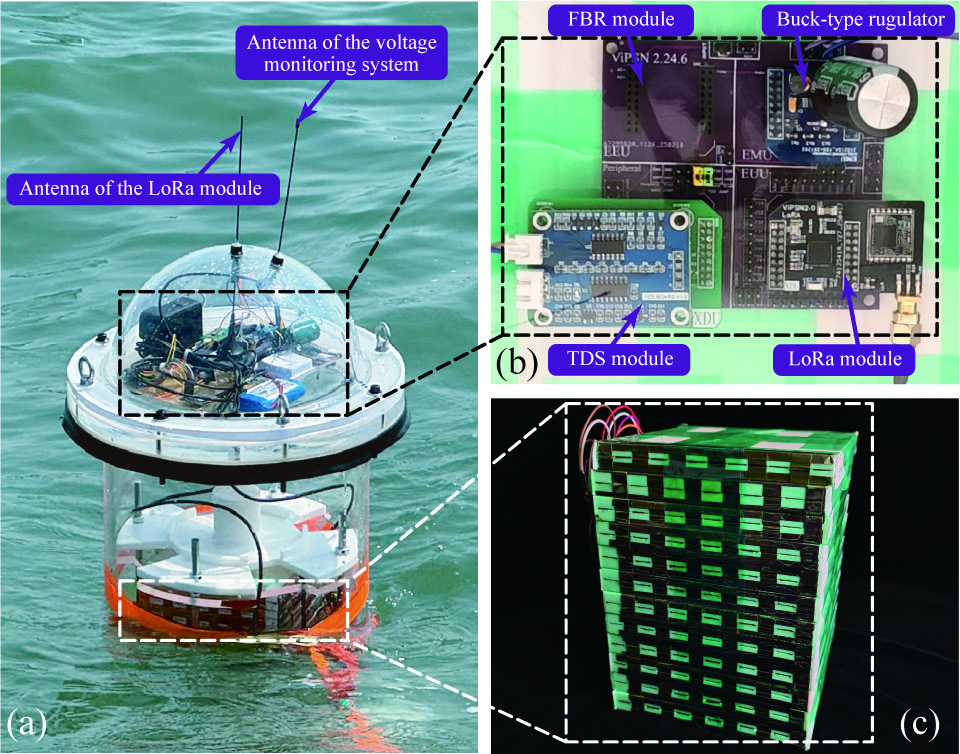}
    \caption{System implementation and field deployment of the ViPSN-LoRa.
    (a) Deployment of a battery-free ocean monitoring buoy. 
    (b) Setup modules of the ViPSN-LoRa, including the FBR module, buck-type regulator, LoRa module, and TDS peripheral for ocean water quality monitoring.
    (c) Internal structure of the TENG array wave energy harvester. 
    }
    \label{fig:fig_case_lora_teng}
\end{figure}

\begin{table}[htbp]
\small
\centering
\caption{System Configuration and Power Consumption of ViPSN-LoRa}
\label{tab:system_lora_results}
\begin{tabularx}{\linewidth}{p{4.2 cm} X}
\hline
\textbf{Parameter} & \textbf{Value} \\
\hline
\textit{Deployment location} & \textit{Bohai Bay, Dalian, China} \\
Excitation condition & Ocean waves, $\sim$0.4\,m height \\
Harvester & TENG array \\
Average harvester output  & 21.63 mW \\
Harvester open-circuit voltage & $\sim$ 530~V \\
Harvester short-circuit current & $\sim 480~\mu$A \\
\hline
Storage capacitance & $6800~\mu$F \\
\hline
Transmission protocol & LoRa UART, 433\,MHz \\
Transmission power & 14\,dBm \\
Transmission data payload  & 12 bytes \\
\textit{Transmission range}  & \textit{0.8\,km} \\
Transmission energy & 22.4~mJ \\
\hline
TDS sampling energy & 1.46~mJ \\
\textit{Sampling interval} & \textit{200--550\,s} \\
\hline
Total energy consumption & 23.86~mJ \\
\hline
\end{tabularx}
\end{table}

\begin{figure}[t]
    \centering
    \includegraphics[width=1\linewidth]{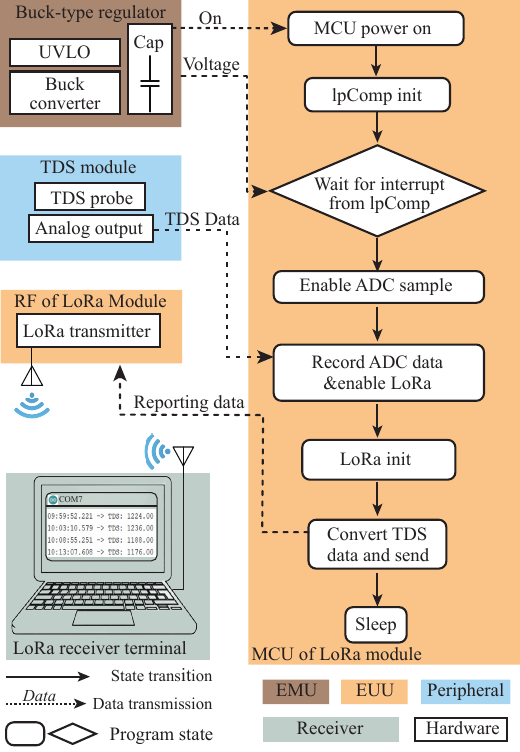}
    \caption{State machine of the ViPSN-LoRa for long-range wireless transmission under intermittent energy supply.}
    \label{fig:fig_case_lora_machine_state}
\end{figure}

\begin{figure*}[t]
    \centering
    \includegraphics[width=2\columnwidth]{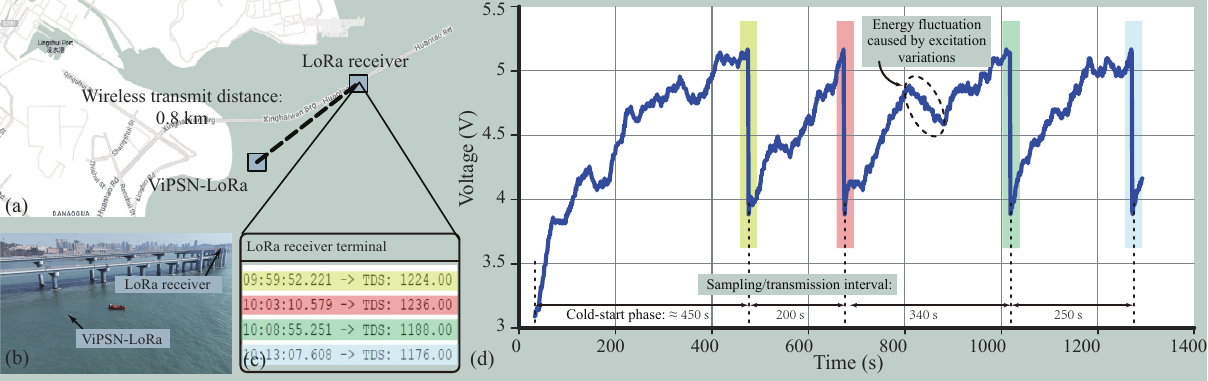}
    \caption{End-to-end deployment and performance assessment of the ViPSN-LoRa.
    (a) Field deployment map in Bohai Bay.
    (b) Ocean buoy equipped with ViPSN-LoRa and the LoRa receiver.
    (c) Real-time sensor data received at the receiver terminal.
    (d) Storage capacitor voltage profile during field testing.}
    \label{fig:fig_case_lora_test}
\end{figure*}

\section{ViPSN-LoRa: Wave-powered TENG LoRa}
This section details the design and field deployment of ViPSN-LoRa, a battery-free marine sensing system built on the ViPSN~2.0 platform. Specifically, we aim to demonstrate the platform's capability to perform heavy-duty, long-range wireless sensing tasks in real-world oceanic environments, leveraging a modular rolling-mode TENG array for sustained wave energy harvesting and robust power management.

\subsection{Setup}
Fig. \ref{fig:fig_case_lora_teng} illustrates the overall system architecture and field deployment setup of ViPSN-LoRa. The prototype, encapsulated within a buoy-type enclosure (Fig. \ref{fig:fig_case_lora_teng}(a)), integrates a modular TENG energy harvesting unit and the ViPSN2.0 hardware modules. Specifically, the system comprises four main functional modules (Fig. \ref{fig:fig_case_lora_teng}(b)):
\begin{itemize}
    \item ETU: TENG array + FBR;
    \item EMU: Buck-Type Regulator with the PID power solution;
    \item EUU: LoRa Module;
    \item Peripheral: TDS sensor for real-time water quality monitoring.
\end{itemize}

The ETU incorporates a planar array of modular TENG harvesters, each securely encapsulated between two acrylic plates at the base of the buoy (Fig.~\ref{fig:fig_case_lora_teng}(c)). This rolling-mode configuration enables energy harvesting from low-frequency, irregular ocean waves\cite{wang2024rolling}. A FBR module converts the alternating output of the TENG array into a unidirectional current suitable for storage. 

The EMU adopts a buck-type regulator implementing the PID solution, featuring a 6800~$\mu$F electrolytic capacitor for energy buffering. The PID thresholds ($V_{\text{Pstart}}=4.7~\text{V}$, $V_{\text{Pgood}} = 5.2~\text{V}$, $V_{\text{Pclose}}$ = $V_{\text{Psleep}} = 3.7~\text{V}$) are optimized to maximize harvested energy utilization while ensuring reliable operation of high-power tasks such as wireless communication. The $V_{\text{Psleep}}$ threshold was omitted in this experiment, as data retention during sleep was not required.

The LoRa module is configured for long-range wireless communication at 433~MHz and 14~dBm transmission power, ensuring reliable data transmission over significant distances. The peripheral unit integrates a TDS module, enabling real-time monitoring of ocean water quality parameters. Given the small size of each water quality sample, the data are packaged into a 12-bytes payload for uplink transmission. Sensor data are periodically sampled and transmitted via the LoRa link to a remote receiver located onshore, demonstrating the platform's practical efficacy in realistic marine IoT applications.

\subsection{Long-Range Transmission}
The state machine of ViPSN-LoRa, as illustrated in Fig.~\ref{fig:fig_case_lora_machine_state}, coordinates the operation of the buck-type regulator, LoRa, and TDS modules. Initially, the system remains disabled until the capacitor voltage reaches the internal startup threshold, at which point the regulator powers up the system. During this low-power waiting state, only the voltage comparator remains active, while the LoRa and TDS modules stay in deep-sleep mode to minimize energy consumption during the energy build-up phase. Once the capacitor voltage exceeds the externally defined threshold, the comparator asserts the $V_{\text{Pgood}}$ signal, waking the system to perform its primary tasks: sampling the TDS module via the ADC, processing the acquired data, and transmitting the results via the LoRa module at 14~dBm for long-range communication.

To meet the heavy-duty demands of LoRa transmissions, the PID solution dynamically adjusts the energy accumulation threshold, ensuring adequate energy is stored before each high power consumption operation. Compared to the fixed-threshold UVLO solution, this solution significantly enhances flexibility and efficiency, enabling ViPSN-LoRa to reliably complete energy-intensive sensing and kilometer-scale transmissions under highly variable energy harvesting conditions.

\subsection{Evaluation}
As shown in Fig.~\ref{fig:fig_case_lora_test}(a) and
(b), we deployed the ViPSN-LoRa buoy in Bohai Bay, Dalian, China, to evaluate its real-world performance and robustness. 
Throughout the field experiment, a wireless voltage monitoring system continuously tracked the energy storage capacitor voltage in real-time, with its antenna explicitly indicated in Fig.~\ref{fig:fig_case_lora_test}(a). 

Fig. ~\ref{fig:fig_case_lora_test}(d) presents the measured voltage waveform, which clearly reveals four distinct voltage drops. Each drop corresponds to a cycle of TDS water quality measurement and a subsequent LoRa data transmission. These transmission events were triggered precisely when the capacitor voltage reached the predefined PID threshold, and concluded when the voltage dropped to the system shutdown threshold.

Experimental results indicate that the initial cold-start charging phase required approximately 450~s, reflecting the time needed for energy accumulation from a fully discharged state. However, subsequent recharging cycles benefited from residual voltage maintained between $V_{\text{Pgood}}$ and $V_{\text{Pclose}}$, significantly reducing recharge time to approximately 200~s per cycle—a reduction of roughly 350~s compared to the initial cold-start cycle.

To validate practical communication performance, we deployed the LoRa receiver approximately 0.8~km onshore, as indicated in Fig.~\ref{fig:fig_case_lora_test}(c). Throughout the experimental period, the sensor data packets transmitted by the buoy were reliably received at the remote terminal, confirming reliable long-range wireless communication. Furthermore, the measured system latency—from reaching the $V_{\text{Pgood}}$ threshold to completing data transmission—remained consistently within a few seconds, demonstrating prompt responsiveness suitable for marine IoT applications.

\begin{figure}[htbp]
    \centering
    \includegraphics[width=1\linewidth]{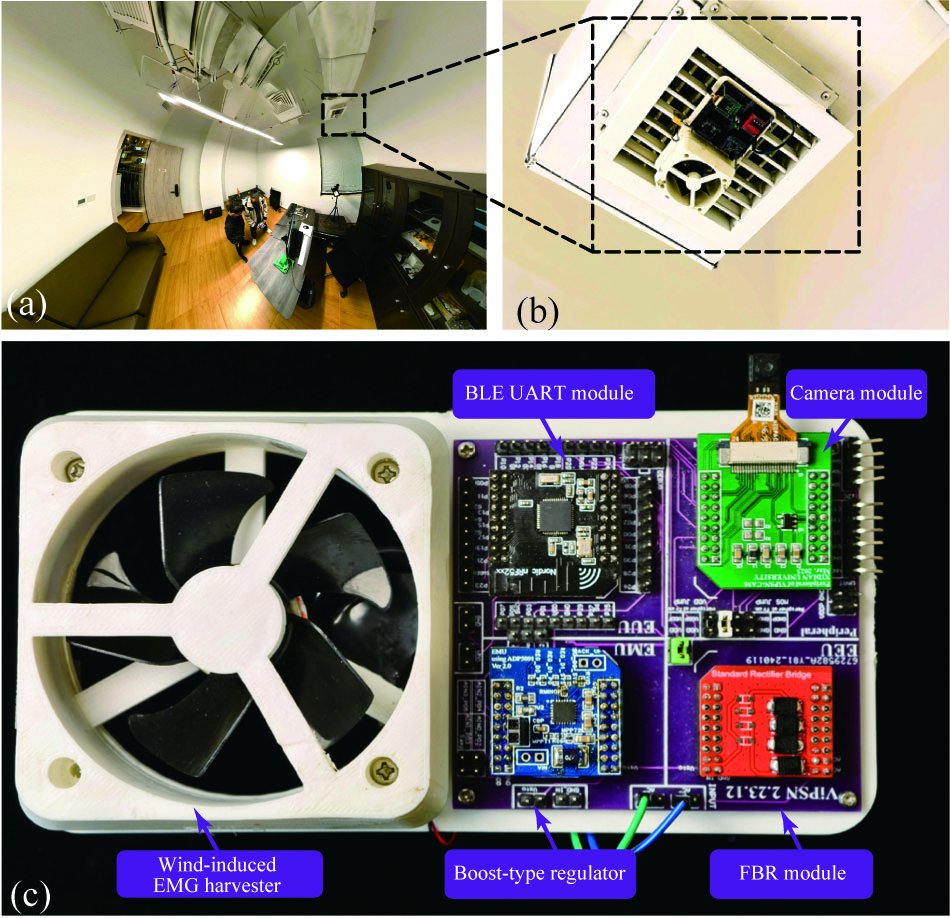}
    \caption{Deployment and configuration of the ViPSN-Cam.
    (a) Experimental deployment environment: a 20~m$^2$ single-person office.
    (b) Close-up view of the camera mounted on an HVAC supply outlet.
    (c) Setup modules of ViPSN-Cam, including a wind-induced EMG harvester, FBR module, boost-type regulator, BLE UART module, camera peripheral, and a carrier board with FRAM for state retention.}
    \label{fig:solution_and_modules}
\end{figure}

\begin{table}[htbp]
\small
\centering
\caption{System Configuration and Power Consumption of ViPSN-Cam}
\label{tab:system_camera_results}
\begin{tabularx}{\linewidth}{p{4.2cm} X}
\hline
\textbf{Parameter} & \textbf{Value} \\
\hline
Deployment location & HVAC ventilation duct \\
Wind speed & 3.5--4.5~m/s \\
Harvester & Wind-induced EMG harvester \\
Average harvester output  & 1.2~mW\\
Harvester open-circuit voltage & $\sim$ 13~V \\
Harvester short-circuit current & $\sim$ 10~mA \\
\hline
Storage capacitance & $4700~\mu$F \\
\hline
Transmission protocol & BLE UART, 2.4~GHz \\
Transmission power & 0~dbm \\
Transmission data payload  & 240~Bytes \\
Transmission energy & 11~mJ \\
\hline
\textit{Image transmission time$^{*}$} & $\sim$\textit{4~s} \\
\textit{Image capture energy} & \textit{16~mJ} \\
\hline
Total energy consumption per image delivery & 27~mJ \\ 
\hline
\end{tabularx}
\vspace{1mm}
\parbox{\linewidth}{\footnotesize%
$^{*}$~Includes the cumulative transmission time of all BLE packets and ADC sampling delays between packet transmissions.}
\end{table}

\begin{figure*}[t]
    \centering
    \includegraphics[width=\linewidth]{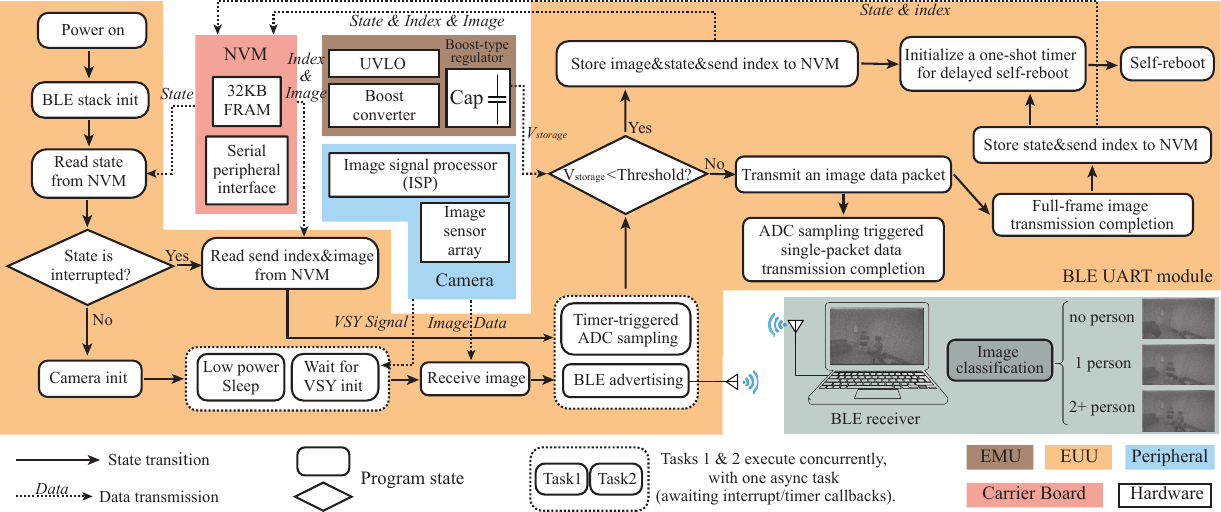}
    \caption{Task-oriented state machine and software architecture enabling intermittent image streaming in the ViPSN-Cam.}
    \label{fig:fsm}
\end{figure*}

\section{ViPSN-Cam: Vibration-powered EMG Camera}
To validate the performance of the APC solution and extend the range of supported applications for ViPSN~2.0, we developed ViPSN-Cam—a battery-free camera system powered by indoor wind-induced vibrations. This prototype is designed to intermittently execute complex-duty tasks such as image capture and wireless data transmission, highlighting the platform’s ability to support demanding, data-intensive IoT workloads in energy harvesting environments.

\subsection{Setup}
The overall system architecture and experimental deployment scenario of ViPSN-Cam are illustrated clearly in Fig.~\ref{fig:solution_and_modules}. As depicted in Fig.~\ref{fig:solution_and_modules}(c), the prototype comprises five main components:

\begin{itemize}
    \item ETU: EMG wind-induced harvester + FBR;
    \item EMU: Boost-type regulator with the APC power solution;
    \item EUU: BLE UART module;
    \item Peripheral: Camera module for image capture;
    \item Carrier Board: FRAM for state retention.
\end{itemize}

The ETU employs an EMG harvester specifically designed to convert indoor airflow-induced vibrations (such as those found inside ventilation ducts) into electrical energy, typically generating approximately 1.2~mW at an airflow speed of 4~m/s.  The EMU employs the APC power solution with actively controlled voltage thresholds ($V_{\text{Pstart}} = V_{\text{Pgood}} = 4.7~\text{V}$, $V_{\text{Pclose}} = 2.2~\text{V}$, $V_{\text{Psleep}} \approx 2.4~\text{V}$ ), optimized to support complex-duty task such as image capture and wireless transmission. Compared to fixed-threshold schemes like PID, APC provides  greater flexibility for complex workloads in both design-time and runtime tuning of key power management parameters. This adaptability is particularly beneficial in avoiding task interruptions during critical operations, making APC well-suited for intermittently powered systems with demanding and time-sensitive computational requirements. The EUU integrates a BLE UART module, coordinating image acquisition, processing, and subsequent wireless data transfer to a remote receiver. The peripheral camera module periodically captures images, while the FRAM module on the carrier board ensures continuous retention of essential system states during periods of insufficient harvested energy.

\begin{figure*}[t]
    \centering
    \includegraphics[width=\linewidth]{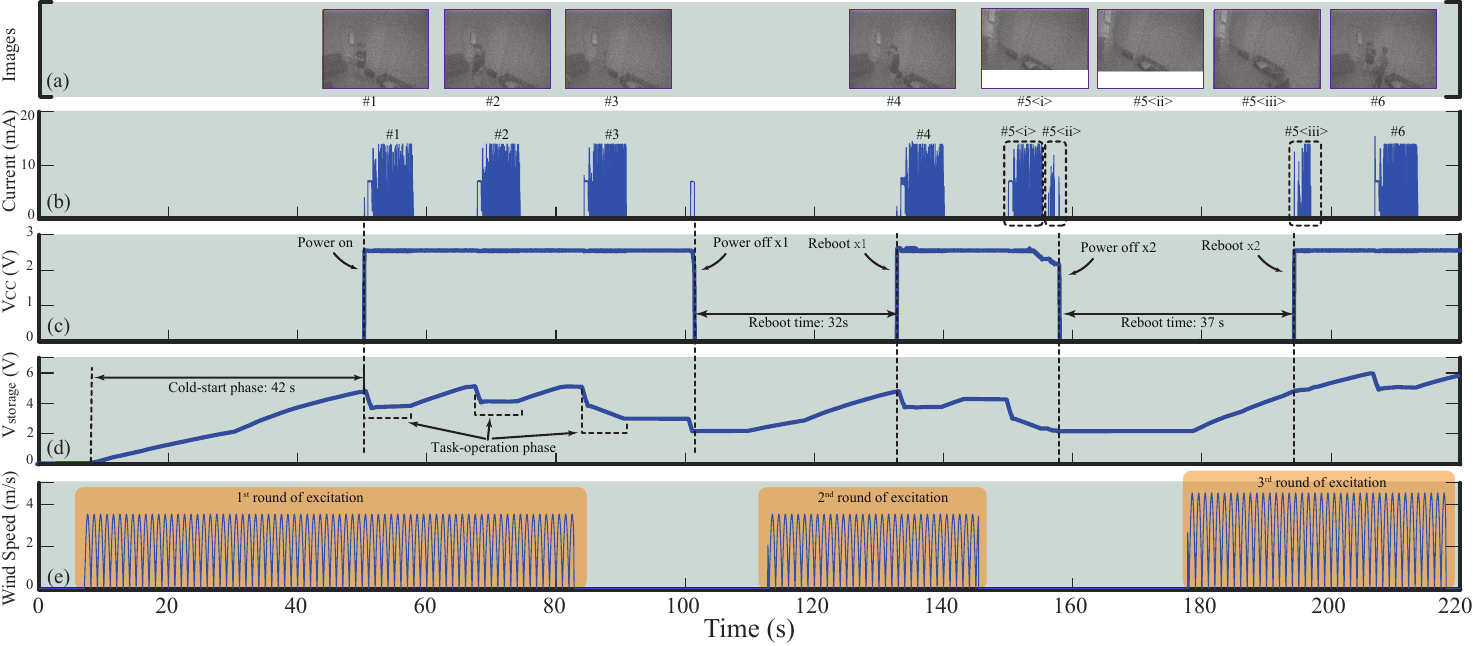}
    \caption{Experimental characterization of the ViPSN-Cam.
    (a) Reconstructed images at the receiver.
    (b) Current consumption of the energy-autonomous camera.
    (c) System supply voltage ($V_{\mathrm{cc}}$).
    (d) Storage capacitor voltage dynamics ($V_\text{storage}$).
    (e) Wind excitation from the HVAC supply outlet.
    }
    \label{fig:cur_vdd_vsto_wind}
\end{figure*}

\subsection{Intermittent Image Streaming}
By leveraging APC-defined voltage thresholds, the system employs a task-oriented state machine to coordinate data preservation and adaptive self-reboot, ensuring reliable intermittent image streaming despite fluctuating energy availability.

\subsubsection{Task-Oriented State Machine}
As shown in Fig. \ref{fig:fsm}, the system operation is controlled by a state machine that governs the execution sequence of tasks. The process is as follows:
\begin{itemize}
    \item Program State Restoration: Upon system wake-up, triggered by $\mathrm{P_{start}}$, the system restores its state from FRAM.
    \item Task Selection: The system determines whether to retrieve unsent data or initializes the camera to capture a new image base on the state restored from FRAM.
    \item Data Acquisition and Voltage Monitoring: During the task-operation phase (between $\mathrm{P_{good}}$ and $\mathrm{P_{sleep}}$), the system acquires data for transmission and periodically samples the energy state.
    \item Conditional Branching:
    \begin{itemize}
        \item If the voltage exceeds the $V_\text{Psleep}$ threshold, the system establishes a BLE connection and begins data transmission.
        \item If the voltage is at or below the $V_\text{Psleep}$ threshold, the system enters the checkpoint phase to save critical data to FRAM and initiates a reboot.
    \end{itemize}
\end{itemize}

Tasks within the dashed boxes in Fig. \ref{fig:fsm} are executed concurrently, including asynchronous operations such as timer callbacks and interrupt handling. This design reduces idle time caused by I/O operations or BLE connection delays, improving energy efficiency.

\subsubsection{Voltage-Triggered Preservation}
To ensure data integrity across power interruptions, the system checks the energy state signaled by $\mathrm{P_{sleep}}$ upon task completion. During BLE broadcasting, the energy state is sampled periodically, and during image data transmission, it is evaluated after each packet is sent. This monitoring enables the system to seamlessly resume transmission from the point of interruption in subsequent power cycles.

\subsubsection{Self-Reboot Mechanism}
Following data preservation during the checkpoint phase, the system triggers a self-reboot, which serves the following purposes:
\begin{itemize}
    \item Accelerated Shutdown: The self-reboot mechanism shortens the transition time to the power-off state, enabling faster restarts in subsequent cycles.
    \item Reduced \(V_\text{Psleep}\) Threshold Precision Requirement: In systems with high energy storage capacity, such as those using large capacitors for camera operation, precise configuration of the \(V_\text{Psleep}\) threshold becomes challenging to implement. When the capacitor voltage falls below \(V_\text{Psleep}\), the system utilizes the remaining energy for further operations until the voltage drops to \(V_\text{Pclose}\), at which point the system powers down.
\end{itemize}

\subsubsection{Data Reception and Classification}
A BLE receiver connected to a laptop computer receives image data transmitted intermittently by the ViPSN-Cam. The received images are processed by a MobileNetV2-based neural network \cite{sandler2018mobilenetv2} running on the computer. The network classifies the images into three predefined categories: ``no person'' (denoting an empty workspace), ``one person'' (indicating a person working at a desk), and ``multiple people'' (representing a meeting scenario), as shown in Fig.~\ref{fig:fsm}. 
This setup demonstrates the capability of the ViPSN-Cam to transmit image data suitable for real-time classification tasks, enabling lightweight computer vision applications such as occupancy detection, object recognition, and scene analysis. It also provides valuable data for energy-efficient building management and occupancy-based control systems.

\begin{figure}[t]
    \centering
    \includegraphics[width=1\linewidth]{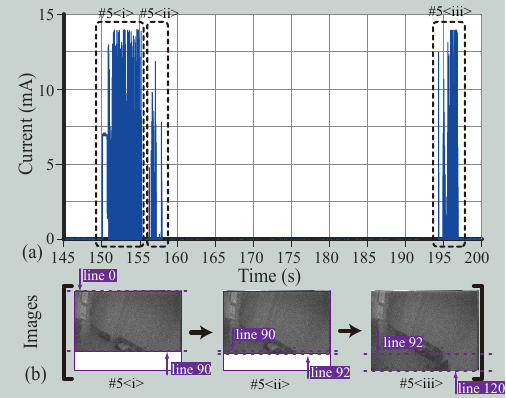}
    \caption{Experimental data capture of image transmission by the ViPSN-Cam during auto-recovery intermittent operation:
    (a) Current consumption profile of the system during intermittent full-image transmission cycles;
    (b) Progressive reconstruction of intermittently transmitted images at the receiver.}
    \label{fig:capture_part}
\end{figure}

\begin{figure}[t]
    \centering
    \includegraphics[width=1\linewidth]{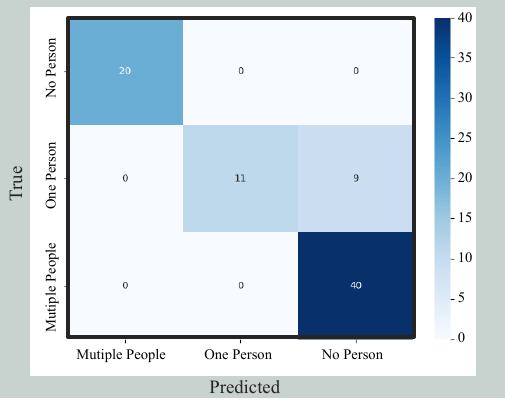}
    \caption{Confusion matrix for the ViPSN-Cam classification task distinguishing between no person, one person, and multiple people scenarios.}
    \label{fig:confusion_matrix}
\end{figure}

\subsection{Evaluation}
To validate the feasibility of battery-free intermittent imaging in realistic ambient conditions, we deployed the ViPSN-Cam prototype beneath an office HVAC supply outlet, as illustrated in Fig.~\ref{fig:solution_and_modules}(a) and (b). 

\subsubsection{Image Capture}
Fig.~\ref{fig:cur_vdd_vsto_wind} illustrates the operational sequence of the system, from cold-start to multiple intermittent image captures under different wind speeds. During the cold-start phase, the harvester charges the storage capacitor from $0~\mathrm{V}$ to the startup threshold voltage of $V_\mathrm{Pstart} = 4.7~\mathrm{V}$ within approximately 42 s. 
With optimized firmware, the system requires $27~\mathrm{mJ}$ to complete a single image capture and data transmission cycle, of which $16~\mathrm{mJ}$ is used for image capture, and $11~\mathrm{mJ}$ is used for BLE connection establishment and data transmission, as shown in Table~\ref{tab:system_camera_results}.

\subsubsection{Intermittent Streaming}
Fig.~\ref{fig:cur_vdd_vsto_wind} provides a detailed illustration of the system's operational resilience through six image capture tasks amidst power interruptions. It comprises five subplots: (a) captured images, (b) instantaneous current consumption, (c) EUU supply voltage ($V_{\mathrm{cc}}$), (d) storage capacitor voltage ($V_\text{storage}$), and (e) ambient wind speed.

The experiment commenced with the ``1st round of excitation'' at 3.5~m/s, during which the system successfully completed a cold-start phase, charging the storage capacitor from 0~V to 4.7~V. Subsequently, it executed 3 consecutive image capture tasks (\#1, \#2, \#3). Following the completion of image \#3, a sudden drop in ambient wind speed caused the storage capacitor voltage to drop below $V_{\text{Pclose}}$, leading to the first power interruption. As this interruption occurred prior to the initiation of the next image capture, the system, upon power restoration via the ``2nd round of excitation'' at 3.5~m/s, recovered its previous operational state from the non-volatile FRAM and proceeded to resume the pending image capture task.

The system then continued its operations, executing image \#4 and initiating image \#5. However, during the BLE data transmission phase associated with image \#5, another sudden drop in ambient wind speed resulted in the second power interruption. Due to the nature of this interruption occurring mid-transmission, the system, upon power restoration via the ``3rd round of excitation'' at 4.5~m/s, continued attempting to transmit the remaining image data. Consequently, as detailed in Fig.~\ref{fig:capture_part}, the complete capture and transmission of image \#5 was achieved in 3 distinct segments.

The second interruption occurred during the transmission of image \#5. Fig.~\ref{fig:capture_part} details the corresponding current profile and image reception progress. Notably, energy usage is concentrated during data transmission, as image streaming over BLE typically requires approximately 4 s, consuming nearly 80\% of the system’s operation time per cycle. As shown in Fig.~\ref{fig:capture_part} \#5\textit{\textless i\textgreater}, the system successfully captured the image and transmitted partial data up to line~90. Triggered by the $P_\text{sleep}$ signal, the system checkpointed its state. Since the capacitor voltage had not yet fallen below $V_\text{Pclose}$, the system resumed operation and transmitted two additional packets, advancing image reception to line~92 (Fig.~\ref{fig:capture_part} \#5\textit{\textless ii\textgreater}). Subsequently, the system performed another checkpoint and entered the shutdown state. Upon the next energy availability and reboot, the system restored its previous state and completed the remaining image transmission, as shown in Fig.~\ref{fig:capture_part} \#5\textit{\textless iii\textgreater}.

\subsubsection{Task Classification}
A total of 130 labeled images were employed for training, with 40 images in each of the first two categories and 50 in the third. Additionally, 80 images were gathered as a separate test set. As shown in Fig.~\ref{fig:confusion_matrix}, the classification model attained an overall accuracy of 88.75\%, with errors mainly resulting from the misclassification of images featuring one person as those with multiple people. Owing to the inherent complexity of discriminating between these two categories under low-energy constraints. In contrast, the model achieved a 100\% classification accuracy for images in the ``no person'' and ``multiple people'' categories, where the visual differences are more pronounced. 
These results demonstrate that the images captured and transmitted by the vibration-powered camera prototype are of sufficient quality to support real-time computer vision tasks, such as image classification.

\section{Conclusion}
In this paper, we presented ViPSN~2.0, a modular and reconfigurable IoT platform specifically designed for vibration-powered, battery-free IoT systems. ViPSN~2.0 integrates multiple vibration transducer technologies—including PZT, EMG, and TENG harvesters—via standardized, interchangeable hardware interfaces, enabling rapid prototyping and deployment across diverse vibration-rich applications.

To systematically address the inherent challenges of intermittent and limited harvested energy, ViPSN2.0 standardizes energy signals and clearly defines their corresponding operational phases at the power solution layer. The flexibility and effectiveness of this approach were validated through three representative case studies: ViPSN-Beacon, demonstrating ultra-low-power wireless transmissions powered by transient mechanical excitations; ViPSN-LoRa, enabling reliable long-range wireless communication using ambient wave-induced vibrations; and ViPSN-Cam, supporting intermittent high-power image capturing and wireless streaming under realistic indoor airflow conditions.

Experimental evaluations confirm that ViPSN~2.0 effectively accommodates a broad spectrum of IoT workloads—from lightweight, periodic sensing to high-power, data-intensive streaming tasks—under practical energy harvesting conditions. This integrated modular design and adaptive energy management approach significantly extend the feasibility and applicability of battery-free IoT systems, providing researchers and developers with a reliable and versatile platform for sustainable IoT deployments in remote, off-grid, and maintenance-challenging environments.

\ifCLASSOPTIONcaptionsoff
  \newpage
\fi

\bibliographystyle{myIEEEtran}
\bibliography{ref}




\end{document}